\documentclass{iopart}
\usepackage{graphics}
\usepackage{verbatim}

\newcommand{\bb}[1]{\mbox{\boldmath${#1}$}}
\newcommand{\lessim}{\;\makebox[0pt][l]{\raisebox{-0.1ex}{${}^<$}}%
  \mbox{\raisebox{0.1ex}{${}_\sim$}}\;}
\newcommand{\gtrsim}{\;\makebox[0pt][l]{\raisebox{-0.1ex}{${}^>$}}%
  \mbox{\raisebox{0.1ex}{${}_\sim$}}\;}

\begin{document}

\title[Atmospheric Newtonian Noise in 3rd-Gen.\ Gravitational-Wave Detectors]{Atmospheric Newtonian Noise May Constrain Third-Generation
  Gravitational-Wave Detectors}

\author{Wenhui Wang and Teviet Creighton}
\address{The University of Texas Rio Grande Valley}
\ead{teviet.creighton@utrgv.edu}

\date{\today}

\begin{abstract}
  Advanced gravitational-wave detector designs are pushing towards
  lower frequencies, where certain types of noise, previously
  considered negligible, may come to dominate the detectors' noise
  budgets.  In particular, we revisit atmospheric Newtonian noise,
  caused by the fluctuating gravitational field as regions of high and
  low gas density move past the detector.  We consider density
  perturbations both due to pressure waves (infrasound) and due to
  advected temperature fluctuations.  In the absence of detailed
  site-specific models of topography, airflow, and building design, we
  present general scaling formulae that estimate the spectrum of
  atmospheric Newtonian noise, and show how it is affected by broad
  detector design choices.  We confirm previous analyses that show
  that atmospheric Newtonian noise is not likely a strong contributor
  to advanced LIGO; however, it will be a dominant factor for
  third-generation detectors targeting low frequencies, with the
  low-frequency cutoff principally constrained by the depth of the
  detector underground.
\end{abstract}


\section{Introduction}
\label{s:introduction}

In 2015, the first direct detection of gravitational waves was made by
the Laser Interferometer Gravitational-wave Observatory
(LIGO)~\cite{lsc:2016}.  Since then, LIGO and similar ``second
generation'' detectors Virgo~\cite{virgo:2015} and
KAGRA~\cite{kagra:2021} have provided a steady stream of gravitational
wave observations.  Now, with the field of gravitational-wave
astronomy firmly established, plans are afoot in the coming decade to
build ``third generation'' detectors, such as the Einstein Telescope
(ET)~\cite{et:2010,et:2020} and Cosmic
Explorer~\cite{reitze:2019,evans:2021}.  At the same time, new
detector concepts such as the proposed Torsion Bar Antenna
(TOBA)~\cite{toba:2020} and Superconducting Omnidirectional
Gravitational Radiation Observatory (SOGRO)~\cite{sogro:2016,sogro:2024} seek to
achieve similar or better sensitivities using novel technologies,
while ambitions space-based detector proposals such as
LISA~\cite{lisa:2017}, DECIGO~\cite{decigo:2019}, and
TianQin~\cite{tianqin:2016} benefit from immense interferometer
baselines and isolation from terrestrial noise sources.

The scientific gains of these proposed detectors come from improved
sensitivity and/or greater spectral range, particularly at low
frequencies.  In doing so, they will encounter new sources of noise
previously deemed unimportant for second-generation detectors.  A
particularly insidious source of noise is Newtonian Noise (NN), the
changing near-zone gravitational field caused by masses moving near
the detector (distinct from the far-zone propagating gravitational
waves from astrophysical objects that these observatories seek to
detect).  Newtonian noise can arise from any sort of mass motion,
including in the ground~\cite{hughes:1998}, the
air~\cite{creighton:2008}, and from human activity~\cite{thorne:1999}.
These sources pose a particular challenge at low
($\lessim10\mathrm{Hz}$) frequencies.  Since detectors cannot be
isolated or shielded from gravitational fields, the only way to
overcome Newtonian noise is to suppress all movement of masses in the
vicinity of the detector, or to model these masses and subtract their
contribution.

In \cite{creighton:2008} one of us (Creighton) derived formulae and
made estimates of atmospheric Newtonian gravitational noise (there
called ``gravity gradient noise'') due to propagating pressure waves
(infrasound) and to thermal variations in the atmosphere being carried
past the detector.  It was motivated primarily by the target
sensitivity of the advanced LIGO detectors, and concluded that these
noise sources would not likely dominate the advanced LIGO noise budget
at 10\,Hz.  Now that advanced LIGO is operating and third generation
observatories are in the planning stages, it seems prudent to revisit
the question of atmospheric gravitational noise from a broader
perspective, considering a wider variety of detector configurations.

This paper presents general scaling laws for the strain noise power
spectral density $S_h(f)$, with tunable parameters to accommodate a
variety of instrumental designs.  In particular, the results will be
applicable to gravity gradiometers (instruments with very short
baselines) as well as long-baseline laser interferometric detectors,
and will consider the effects of moving the instrument underground or
even (in principle) near-Earth space.  Our goal is not, primarily, to
give specific predictions for any particular design, but to provide a
more general analytic tool to evaluate and assess noise ``risk'' for
different design concepts.

The paper is organized as follows: Section~\ref{s:infrasound} gives
estimates of the noise due to low-frequency pressure perturbations
(infrasound), and demonstrates the broad comparability of our
formalism with the results of other researchers~\cite{fiorucci:2018}.
Section~\ref{s:thermal} estimates the noise due to temperature
perturbations advected past the instrument, which represents a more
concerning and difficult-to-mitigate source of Newtonian noise.
Section~\ref{s:cases} presents some typical results for certain
canonical design choices.  Section~\ref{s:discussion} discusses the
general prospects for advanced detectors, concluding with the
observation that these detectors' low-frequency response will be
largely constrained by their depth underground.  An appendix presents
numerical simulations that were undertaken to check the validity of
our analytic formulae.

\section{Infrasonic gravitational noise}
\label{s:infrasound}

Infrasound consists of atmospheric pressure fluctuations below 20\,Hz.
Infrasound researchers~\cite{matoza:2013} distinguish between coherent
infrasound (pressure waves propagating from remote sources) and
incoherent or wind-generated infrasound (pressure changes caused by
turbulent airflow at the receiver).  For purposes of gravitational
noise, we are concerned primarily with coherent infrasound;
wind-driven infrasound would behave in a manner similar to thermal
fluctuations described in section~\ref{s:thermal}, but with density
perturbations about 4 orders of magnitude lower.

We start with a derivation of the gravitational perturbation due to a
coherent plane pressure wave, in a slightly more generic form than
in~\cite{creighton:2008}.  We will first give the general scaling law,
then the more precise form of the gravitational noise for a coherent
plane wave, then show approximately how various real-world effects
modify the ideal form.

The gravitational field of an infinite plane of density $\rho$ and
thickness $w$ is $g=2\pi\rho w$.  In a plane wave train, successive
planes of density perturbation $\pm\delta\rho$ largely cancel each
other out, but this cancellation is not exact due to incoherence of the
wave train.  The result is that the effective field of the entire
train comes from the asymmetry of the density perturbation
$\pm\delta\rho$ at a distance $\pm\lambda/2$ on either side of a given
point, giving a gravitational field:

\begin{equation}
  \label{eq:dg-plane-approx}
  \delta g \;\;\sim\;\; G\delta\rho\,\lambda
\end{equation}

\noindent More precisely, consider a wave train of density
perturbations $\rho=\rho_0+\rho_1\cos(\omega t-kx)$, with
$k=2\pi/\lambda=\omega/v_s$ where $v_s$ is the sound speed.  The
disturbance in the gravitational field is given by:
$$\begin{array}{rcl}
  \delta g_x &=& \displaystyle
  \int_{-\infty}^\infty 2\pi G\rho_1\cos(\omega t-kx)\,\mathrm{sgn}(x)\,dx \\[2ex]
  &=& \displaystyle2\pi G\rho_1\left(\int_0^\infty\cos(\omega t-kx)\,dx -
  \int_0^\infty\cos(\omega t+kx)\,dx\right)
\end{array}$$
Although not formally integrable, we can solve this using a
regularizing factor $e^{-ax}$ with $a\ll k$ (physically equivalent to
imposing incoherence on a scale $l\gg\lambda$):

\begin{equation}
  \label{eq:dg-plane}
  \delta g_x \;\;=\;\; 2\pi G\rho_1\sin\omega t\,\mbox{$\frac{2}{k}$}
  \;\;=\;\; 2\pi G\delta\rho\,\mbox{$\frac{2}{k}$} \;.
\end{equation}

For an interferometer, the effect of this acceleration is to produce
an effective strain $h(t)$ on an arm of length $L$ given by:
$$
\ddot{h} \;\;=\;\; \frac{\delta g}{L}
\;\;=\;\; 4\pi G\delta\rho\,\frac{v_s}{\omega L}
$$
The noise from the four test masses at the ends of the two arms will
add in \emph{quadrature}, increasing this by a factor between
$\sqrt{3}$ and 2, depending on correlations between the noise in the
two masses at the corner.  Pessimistically, we assume a factor of 2.

For a gravity gradiometer sensor, the measured gravity gradient is
$g'=kg$, or:
$$
\delta g' \;\;=\;\; 4\pi G\delta\rho
$$
In either case, the contribution to the gravitational wave signal $h$
comes by taking two time integrals of the above quantities,
introducing a factor of $\omega^{-2}$.  The gravitational wave noise
$S_h$ is thus related to the atmospheric density noise $S_\rho$ via
$$
\sqrt{S_h} \;\;\sim\;\; \left\{\begin{array}{l@{\quad}l}
4\pi G\sqrt{S_\rho}\,\frac{1}{\omega^2} & \mbox{gradiometer} \\[1ex]
4\pi G\sqrt{S_\rho}\,\frac{1}{\omega^2}\,\frac{2v_s}{\omega L} &
\mbox{interferometer}
\end{array}\right.
$$
For sound waves, one can relate density fluctuations to pressure
fluctuations via $(\delta P/P)=\gamma(\delta\rho/\rho)$, where
$\gamma\approx1.4$ is the adiabatic index.  Further note that an
interferometer effectively becomes a gradiometer when the arm length
$L$ becomes much less than the wavelength of the perturbation.  So the
above two expressions can be collected together as:
$$
\sqrt{S_h} \;\;\sim\;\; 4\pi G\rho\,\frac{\sqrt{S_P}}{\gamma P}\,
\omega^{-2}\left(1+\frac{\omega L}{2v_s}\right)^{-1}
$$
The derivation in~\cite{creighton:2008} identifies two additional
factors: a factor of $1/2$ due to the fact that the pressure wave
covers only the half-space above the ground, and a geometric factor of
$\sqrt{\langle\cos^2\theta\rangle}$ for the average projection of the
plane wavevector onto an arm axis.  The geometric factor is
$1/\sqrt{3}$ for an isotropic distribution of waves, or $1/\sqrt{2}$
for waves in or near the horizontal plane; we take the realistic (and
pessimistic) value of $1/\sqrt{2}$.

\subsection{Shielding}

The vacuum system and surrounding buildings will cut off the
high-frequency component of gravitational noise, by excluding
small-scale density perturbations in the immediate vicinity of the
test masses.  In~\cite{creighton:2008} this effect was computed for
the case of a cylindrical end station, giving a spectrum with features
specific to that geometry, while at the same time ignoring
certain effects such as reflection and diffraction that would affect
those features.

In this section we will instead consider a plausible but easier to
model scenario where the test mass is located some distance $D$ below
ground.  We will take the result as a reasonable order-of-magnitude estimate for \emph{any} instrument that isolates the test masses some distance $\sim D$ from the noisy environment, without giving undue emphasis to features that
would be specific to a particular geometry.

The field in this
case is:
\begin{eqnarray}
    g_x &=& \displaystyle\int G\rho\,\frac{x}{r^3}\,dV \nonumber\\
    \delta g_x &=& \displaystyle\int_{x=-\infty}^\infty
    \int_{y=-\infty}^\infty \int_{z=D}^\infty G\rho_1\cos(\omega t-kx)\,
    \frac{x\,dz\,dy\,dx}{(x^2+y^2+z^2)^{3/2}}\nonumber\\
    \label{eq:dg-plane-underground}
    &=& 2G\rho_1\sin(\omega t)\int_0^\infty\sin(kx)
    \arctan\left(\frac{x}{D}\right)\,dx
  \end{eqnarray}

\noindent We note that as $D\rightarrow-\infty$,
$\arctan(x/D)\rightarrow\pi\,\mathrm{sgn}(x)$, and we recover the result
in Eq.~(\ref{eq:dg-plane}); whereas when $D\rightarrow0$,
$\arctan(x/D)\rightarrow\frac{\pi}{2}\,\mathrm{sgn}(x)$, or half the
previous result, corresponding to an unshielded test mass on the
surface.  However, the case of interest is when $D$ is positive, in
which case the integral over $x$ gives:
\begin{equation}
\label{eq:gx-shielded}
\delta g_x \;\;=\;\; \pi G\,\delta\rho\,\mbox{$\frac{2}{k}$}\,e^{-kD} \;.
\end{equation}
Including the quadrature sum of four test masses and the projection factor of
$\sqrt{\langle\cos^2\theta\rangle}=1/\sqrt{2}$, we arrive at the
approximate gravitational noise on a shielded test mass:

\begin{equation}
  \label{eq:sh-semiplane}
  \sqrt{S_h} \;\;\approx\;\; \sqrt{2}\pi G\rho\,\frac{\sqrt{S_P}}{\gamma P}\,
  \omega^{-2}\left(1+\frac{\omega L}{2v_s}\right)^{-1}\,e^{-\omega D/v_s} \;.
\end{equation}

Note that we have assumed here a wave propagating horizontally.  This
is motivated by the fact that most coherent outdoor infrasound will be
generated by surface phenomena some distance from the detector.  These
results do generalize, approximately, to waves with some vertical
component.  In that case one has both an incident and reflected wave
train.  While the two wave trains nominally double the gravitational
response $\delta g$, they also double the measured pressure
fluctuation $\delta P$ at ground level (within a fraction of a
wavelength from the surface).  Thus, to a reasonable approximation,
Eq.~\ref{eq:sh-semiplane} is unaffected by ground reflection, provided
we assume $\sqrt{S_P}$ is measured near ground level.  Further
discussion of ground reflection can be found in~\cite{creighton:2008}
sec.~2.1.

\subsection{Incoherence}

The exponential cutoff in Eq.~(\ref{eq:sh-semiplane}) requires some
explanation.  It arises because each wave front in the wavetrain
affects the test mass \emph{coherently} over a timescale $D/v_s$,
which smooths out the gravitational perturbation over that timescale.
A basic principle of Fourier transforms is that smoothing (convolving)
the time series over a timescale $\tau$ truncates (multiplies) the
spectrum on a frequency scale $\omega\sim1/\tau$.

However, for very large $D$, one must consider the fact that the plane
wave is only coherent over scales $l=Q/\lambda$, where $Q$ is
a ``quality factor'' determining the spatial coherence
of the wave, and $l$ is the \emph{coherence length}.  This means that
the exponential cutoff will \emph{not} continue beyond $D\sim l$,
becoming instead a simple power law dependence due to the weakening
gravitational field.

Determination of $l$ or $Q$ is beyond the scope of this paper.  We note that coherence lengths may be only a few wavelengths~\cite{2012AREPS..40..327H}.  It is worth pointing out that, although we are concerned with the time-varying perturbation on the instrument, this is determined by the sound wave's \emph{spatial} coherence: what matters is the duration that \emph{each individual} wavefront smoothly affects the test mass, rather than its correlation with the wavefronts that precede or follow it.  Ultimately this spatial coherence will depend on dissipation, scattering, and distance to the source(s), as opposed to whether the source is temporally coherent.
We further note that for the frequencies of interest, infrasound wavelengths
are many tens of metres, so that even moderate $Q$
values will mean that this effect matters only for deeply-buried
detectors.  However, several detector proposals \emph{do} posit depths
of hundreds of metres.  While incoherence does not seem to be a significant effect for current design proposals (see section~\ref{s:cases}), we include here a derivation of its impact, for the sake of completeness and for possible future detectors.

Consider the contribution of a single ``packet'' of coherent
infrasound with length $l$ in each dimension.  We start by restricting
the wave train to a horizontal tube of area $A\sim l^2$ a distance $D$
above the detector:
$$
\begin{array}{rcl}
  g_x &=& \displaystyle\int G\rho\,\frac{x}{r^3}\,dV
  \;\;=\;\; \int_{-\infty}^\infty G\rho l^2\,\frac{x}{r^3}\,dx
\end{array}
$$
For $D\gg l$, we can treat the tube as effectively a line mass of
linear density $\rho l^2$, giving a gravitational perturbation of:
$$
\begin{array}{rcl}
  \delta g_x &\approx& \displaystyle 2G\rho_1l^2\sin\omega t
  \int_0^\infty\sin(kx)\,\frac{x}{(x^2+D^2)^{3/2}}\,dx \\[2ex]
  &=& 2Gl^2\delta\rho\,kK_0(kD)
\end{array}
$$
where $K_0$ is the modified Bessel function of the second kind of
order 0.

The Bessel function still has an exponential cutoff $\sim e^{-kD}$ for
$kD\gg1$, because we have not yet introduced the constraint in $x$ that
would reduce the temporal smoothing.  Let $A(x)$ be some
to-be-specified wave packet envelope with lengthscale $l$.  Then the
perturbation will be:
$$
\begin{array}{rcl}
  \delta g_x &\approx& \displaystyle 2Gl^2\rho_1l^2\sin\omega t
  \int_0^\infty\sin(kx)A(x)\frac{x}{(x^2+D^2)^{3/2}}\,dx \\[2ex]
  &=& \displaystyle 2Gl^2\delta\rho
  \int_0^\infty\sin(kx)A(x)B(x)\,dx
\end{array}
$$
where $B(x)=x(x^2+D^2)^{-3/2}$.  As noted earlier, the Fourier transform of a product is the
convolution of the transforms, and vice-versa.  Thus while
the Fourier transform $\tilde{B}(k)=kK_0(kD)\sim e^{-kD}$ for large $kD$, it will be convolved
over $k$ by a function $\tilde{A}(k)$ with a much wider support
over $k\sim1/l\gg1/D$, spreading out the exponential cutoff. That is:
$$
\begin{array}{rcl}
  \delta g_x &\approx& Gl^2\delta\rho\sin(\omega t)\;\Im\left\{
  \int_{-\infty}^\infty e^{ikx}A(x)B(x)\,dx\right\}
  \\[2ex]
  &=& Gl^2\delta\rho\sin(\omega t)\;\Im\left\{\frac{1}{2\pi}\,\int_{-\infty}^\infty \tilde{A}(\kappa)\tilde{B}(k-\kappa)\,d\kappa\right\}
\end{array}
$$
For now we do not
specify $A(x)\rightleftharpoons \tilde{A}(k)$, but note that:
$$
\begin{array}{rcl}
  \tilde{B}(k) &=& \int_{-\infty}^\infty e^{ikx}B(x)\,dx \\[2ex]
  &=& \displaystyle 2i\int_0^\infty \sin(kx)\,\frac{x}{(x^2+D^2)^{3/2}}\,dx \\[3ex]
  &=& 2ikK_0(kD)
\end{array}
$$
Now $kK_0(kD)$ is an odd-symmetry function of $k$ with support over
$|k|\sim1/D$.  Using the Bessel identity $\frac{d}{dz}[z^\nu
  K_\nu(z)]=z^\nu K_{\nu-1}(z)$, we can write:
$$
\begin{array}{rcl}
  \tilde{B}(k) &=& -2i\,\frac{1}{D}\,\frac{d}{dk}[kK_1(kD)]
\end{array}
$$
where $kK_1(kD)$ is an even-symmetry function of $k$ with support over
$|k|\sim1/D$.  Moreover, it is absolutely integrable over the real
axis with $\int_{-\infty}^\infty kK_1(kD)\,dx=\pi/2D^2$.  Thus
in the limit of large $D$:
$$
\frac{2D^2}{\pi}\,kK_1(kD) \;\;\sim\;\; \delta(k)\,,\qquad
\frac{2D^3}{\pi}\,kK_0(kD) \;\;\sim\;\; \delta'(k)\,
$$
so long as we are convolving against another function whose support is
$|k|\gg1/D$.  Hence:
\begin{eqnarray}
  \delta g_x &\sim& \displaystyle Gl^2\delta\rho\;\Im\left\{-\frac{i}{2D^3}
  \int_{-\infty}^\infty \tilde{A}(\kappa)\delta'(k-\kappa)\,d\kappa\right\} \nonumber\\[2ex]
  &=& \displaystyle Gl^2\delta\rho\;\Im\left\{\frac{i}{2D^3}\tilde{A}'(k)\right\}
  \;\;=\;\; - G\frac{l^2}{2D^3}\delta\rho\;\Re\left\{\tilde{A}'(k)\right\}
\end{eqnarray}

The choice of an envelope $A(x)$ is somewhat arbitrary in the absence
of detailed site-specific infrasound coherence measurements.  But the
qualitative behaviour does not change much, in the $D\gg l$ limit we
are considering here.  For simplicity we will assume a Lorentzian envelope with full-width half-maximum $l$:
$$
A(x) \;\;=\;\; \frac{(l/2)^2}{x^2+(l/2)^2}
\quad\rightleftharpoons\quad
\tilde{A}(k) \;\;=\;\; \frac{\pi l}{4}\,e^{-kl/2}
$$
This gives the following perturbation for the packet:
\begin{equation}
  \delta g_x \;\;\approx\;\; \pi G\frac{l^4}{16D^3}\,e^{-kl/2}\,\delta\rho\;,\qquad D\gg l
\end{equation}
Qualitatively, we can understand this as follows: The net mass
perturbation of the coherent packet is zero, so we expect a dipole
behaviour $\sim D^{-3}$ at large distance.  The total mass
contributing to the dipole is $l^3\delta\rho$ giving a dipole moment $\sim
l^4\delta\rho$; however this is reduced by a factor $e^{-kl/2}$ due to
the fact that the envelope varies smoothly over a lengthscale $\sim l/2$ (so that each wavefront affects the test mass smoothly over a timescale $\sim kl/2\omega$).

We have validated these analytic results with numerical simulations of a variety of envelope shapes, mostly drawn from families of $C_\infty$ test functions.  Generically all give tails $\propto e^{-\alpha kl}$, though the prefactor and dimensionless $\alpha$ depend on the details of the envelope.  (We note that a Gaussian envelope gives a qualitatively different $l^2$ dependence in the exponent, but this is specific to the Gaussian shape while all others have the generic linear exponential dependence.)

But we are not quite done yet!  Since we are considering the limit
$D\gg l$, there will be a contribution from \emph{many} packets
$N_\mathrm{eff}$ whose effects combine \emph{incoherently:} that is,
they add linearly in power (or quadrature in amplitude).  The enhancement factor replaces the $1/D^3$ for a single packet with:
$$
\begin{array}{rcl}
  \displaystyle \frac{N_\mathrm{eff}}{D^3}
  &=& \displaystyle \sqrt{\raisebox{1ex}{$\displaystyle\sum_\mathrm{packets}$}\bigg(\frac{1}{r^3}\bigg)^{\!\!2}}
  \;\;=\;\; \sqrt{\int\frac{1}{r^6}\,\frac{dV}{l^3}} \\[2.5ex]
  &=& \displaystyle \sqrt{\frac{1}{l^3}\int_{x=-\infty}^\infty\int_{y=-\infty}^\infty\int_{z=D}^{D+H}
    \frac{dz\,dy\,dx}{(x^2+y^2+z^2)^3}} \\[2.5ex]
  &=& \displaystyle \sqrt{\frac{\pi}{6l^3}\left(\frac{1}{D^3}-\frac{1}{(D+H)^3}\right)}
\end{array}
$$
where we introduce a cutoff height $H$ at the top of the atmospheric boundary layer (typically several hundred metres).  The expression in parentheses goes as $1/D^3$ for $D\ll H$ and as $3H/D^4$ for $D\gg H$.  We therefore interpolate among these limits by setting:
$$
\frac{N_\mathrm{eff}}{D^3} \;\;\approx\;\; 
\sqrt{\frac{\pi}{6l^3}\left(D^3+\frac{D^4}{3H}\right)^{-1}}$$
which is numerically safer to compute, has the correct limits, and is order-of-magnitude correct between these limits.  This gives us:
\begin{equation}
  \delta g_x \;\;\approx\;\; \pi G\delta\rho\;e^{-kl}\sqrt{\frac{\pi l^5}{1536}\left(D^3+\frac{D^4}{3H}\right)^{-1}}\;,\qquad D\gg l
\end{equation}
We can combine this with Equation~(\ref{eq:gx-shielded}), which corresponds to the limit $D\ll l$, by adding the two exponential terms (that is, asymptoting to the one that gives the \emph{least} attenuation) and combining the prefactors in inverse quadrature:
\begin{eqnarray}
  \delta g_x &\approx& \pi G\delta\rho\;\frac{2}{k}\left(1+\frac{6}{\pi}\left(\frac{32}{kl}\right)^2\left(\frac{D}{l}\right)^3\left(1+\frac{D}{3H}\right)\right)^{-1/2}\nonumber\\
  &&\qquad\times\left(e^{-kD}+e^{-kl/2}\right)\;,\qquad\mbox{fit for arbitrary $D$}
\end{eqnarray}
Since coherence length will typically scale with wavelength, we write the expression in terms of a \emph{quality factor} $Q=l/\lambda=kl/2\pi=\omega l/2\pi v_s$. 
 Combining with Eq.~(\ref{eq:sh-semiplane}) results in the following
fitting formula covering all the limiting cases:
\begin{eqnarray}
  \sqrt{S_h} &\approx& \sqrt{2}\pi G\rho\,\frac{\sqrt{S_P}}{\gamma P}\,
  \omega^{-2}\left(1+\frac{\omega L}{2v_s}\right)^{-1}
  \left(e^{-\omega D/v_s}+e^{-\pi Q}\right)\nonumber\\
  \label{eq:sh-incoherent}
  &\times&\left(1+\frac{6}{\pi}\left(\frac{2}{\pi Q}\right)^5\left(\frac{\omega D}{v_s}\right)^3\left(1+\frac{D}{3H}\right)\right)^{-1/2}\;.
\end{eqnarray}
This recovers Eq.~(\ref{eq:sh-semiplane}) in the infinite plane wave limit $Q\rightarrow\infty$, but also gives a power-law tail for finite $Q$.  If the measured infrasound is only marginally coherent ($Q\sim1$) this may limit the low-frequency gains achieved by using underground detectors.

\subsection{Surface Structures}
\label{ss:surface-structures}

Buildings on the surface provide less NN shielding, for a given minimum standoff distance $D$, than buried detectors.  Obviously when a detector is surrounded by density perturbations at a distance $D$ on all sides, there will be more mass contributing to the high-frequency short-timescale variations than when this minimum separation is achieved only in a single (upward) direction.

A proper derivation of the effect of surface structures would not only have to limit the volume integral in Eq.~(\ref{eq:dg-plane-underground}), but should also include the effects of reflection, diffraction, and absorption from these structures.  This is outside the purview of our generic treatment here.  Instead, we propose a heuristic estimate of the ``equivalent effective depth'' of a subterranean detector that would provide a similar level of shielding, then use that depth in Eq.~(\ref{eq:sh-incoherent}).  We consider the (approximate) validity of this heuristic in section~\ref{ss:comparison-with-other-published-results}.

Since the frequency cutoff in Eq.~(\ref{eq:sh-incoherent}) goes as $1/D$, our effective distance is based on the angle-averaged inverse distance from the detector to the outside air:
$$
\mbox{$\langle\frac{1}{r}\rangle\;\;=\;\;\frac{1}{4\pi}\int\frac{d\Omega}{r}$}
$$
where $r$ is the distance to the airmass in the direction of the solid angle element $d\Omega$, and we integrate only over directions that reach the outside air (i.e.\ treat all non-intersecting directions as having $r=\infty$).  For a detector a distance $D$ underground we have $\langle\frac{1}{r}\rangle=\frac{1}{4D}$, so for other geometries we can define an ``effective depth''

\begin{equation}
\label{eq:effective-depth}
\mbox{$D_\mathrm{eff}\;\;=\;\;\frac{1}{4}\,\langle\frac{1}{r}\rangle^{-1} \;\;=\;\; \pi\left(\int\frac{d\Omega}{r}\right)^{-1}$}
\end{equation}

\noindent Examples: A detector at the centre of a hemispherical dome of radius $R$ has $D_\mathrm{eff}=\frac{1}{2}R$; while a long semicylinder gives $D_\mathrm{eff}=\frac{2}{\pi}R$.  A typical building might be modeled as a half-cube of height $h$ (side $2h$); for this we numerically evaluate $\langle\frac{1}{r}\rangle=0.4562\frac{1}{h}$ or $D_\mathrm{eff}=0.5480h$, about the same as a hemisphere.  Other geometries can be approximately interpolated among these cases.

\subsection{Infrasound Measurements}
\label{ss:infrasound-measurements}

Extensive measurements of typical infrasound spectra worldwide are now
readily available due to the International Monitoring System (IMS) for
nuclear explosions; see~\cite{woodward:2005} for examples and
summaries of infrasound noise spectra.  However, these spectra are
dominated by incoherent wind-generated infrasound, and do not separate
out the coherent component that is most significant for gravitational
noise.  Infrasound monitoring stations (such as would need to be
deployed at a gravitational-wave detector site) are in fact designed
to minimize incoherent infrasound and discriminate coherent
infrasound; however, the research in this field has focused on the
ability to detect discrete coherent bursts in the incoherent
background, as described for example in~\cite{matoza:2013}, rather
than characterizing the weaker background of \emph{unresolvable}
coherent sources that would dominate the gravitational noise spectrum.

Naturally, to determine the expected infrasonic gravitational noise in
a detector, one would want to deploy an infrasound sensor array
(following design principles similar to an IMS station) at the actual
or proposed site.  Nonetheless one can use the above-mentioned
research to estimate a typical range of expected coherent noise
levels.  The worst case (from the perspective of gravitational-wave
detector noise) would be if all or most of the measured infrasonic
noise were coherent.  In~\cite{matoza:2013} the median RMS noise
level, over the entire year across 39 sites worldwide, decreases
gradually with frequency up to 1\,Hz and is fairly flat above 1\,Hz at
an RMS amplitude of $\sim5\,\mathrm{mPa}$ in bands of width $\Delta\ln
f=0.414=(\ln500)/15$ (that is, 15 bands of equal logarithmic width in
a 500-fold frequency range from 0.01\,Hz to 5\,Hz).  This gives
$\sqrt{S_P}=(8\,\mathrm{mPa}/\sqrt{\mathrm{Hz}})(f/\mathrm{Hz})^{-1/2}$
as an upper limit on the median infrasonic noise.

At the other extreme, in~\cite{matoza:2013}, \emph{coherent} burst
events were resolved with a similar median RMS amplitude
$\sim5\,\mathrm{mPa}$ in the bands above 1\,Hz.  The RMS amplitude
remains fairly flat with frequency, but the time integration window
varies inversely with frequency (with $\Delta t=31\,\mathrm{s}$ for
the band spanning 1\,Hz), implying an average total energy per
recorded burst that decreases with frequency.  The median number of
recorded bursts remained steady at $10^4$ per month in each band above
0.2\,Hz.  Spreading the energy of detected coherent bursts over the
entire month gives a background RMS amplitude of
$5\,\mathrm{mPa}\times\sqrt{10^4\times31\,\mathrm{s}/30\,\mathrm{days}}\sim1.7\,\mathrm{mPa}$
for the 0.5\,Hz wide band around 1\,Hz, or
$\sqrt{S_P}\approx3\,\mathrm{mPa}/\sqrt{\mathrm{Hz}}$.  This is
presumably a \emph{lower} limit, since it neglects any continuous
background of weaker \emph{unresolvable} coherent waves, which are
the source of greatest concern for gravitational noise.

We therefore estimate a \emph{typical} coherent infrasound spectral
density of:
$$
\sqrt{S_P} \;\;\sim\;\; (\mbox{3--8}\,\mathrm{mPa}/\sqrt{\mathrm{Hz}})
\left(\frac{f}{\mathrm{Hz}}\right)^{-1/2}
$$
This is comparable to the
$\sqrt{1000\,\mathrm{nbar}^2/\mathrm{Hz}}\approx3\,\mathrm{mPa}/\sqrt{\mathrm{Hz}}$
baseline assumed in~\cite{creighton:2008}.  Pessimistically, the
$95^\mathrm{th}$ percentile noise levels are at least an order of
magnitude higher than this.

\subsection{Comparison With Other Published Results}
\label{ss:comparison-with-other-published-results}

Other researchers have been active in the area of infrasonic Newtonian noise; we call particular attention to the excellent work by Fiorucci et al.~\cite{fiorucci:2018}.  Direct comparisons can be challenging because the methods, goals, and assumptions behind the two efforts are somewhat different.  The results in in~\cite{fiorucci:2018} are specific numerically-computed outcomes for particular sites and geometries, whereas our intent is to give \emph{analytic} estimates of more general cases, with variables that represent broad design parameters rather than specific dimensions.  (But see Section~\ref{s:cases} where we apply our formula to particular case studies.)  Nevertheless, the results of the two research efforts are essentially consistent, and can be seen as a numerical confirmation of our analytic model.

We note that Eq.~(2) in~\cite{fiorucci:2018} is essentially the same as our Eq.~(\ref{eq:sh-semiplane}), once one makes the following notational changes.  First, their expression has explicit incident and reflected waves at an angle to the surface, and their pressure perturbation $\delta p$ is the amplitude of \emph{each} of these waves.  Our expression is for a horizontally-propagating wave: it is also approximately correct (up to an angle cosine) for oblique waves, if one treats the pressure noise as the \emph{total} (incident plus reflected) measured at or near the surface: our transfer function must be multiplied by 2 to be compared with theirs.  Second, however, our expression has to be \emph{divided} by 2 to get the effect on a single test mass (which is what their formula refers to).  Third, our expression must be multiplied by $L\omega^2$ in the large-$L$ limit to convert from strain to gravitational acceleration, and divided by $k=\omega/v_s$ to convert to gravitational potential $\phi$ (since the inverse process, taking the gradient of the potential, corresponds to multiplying the Fourier transform by $k$).

\begin{figure}
  \begin{center}
    \begin{tabular}{cc}
      \rotatebox{90}{\makebox[2in][c]{$L\omega^2\sqrt{S_h/S_P}$}} &
      \resizebox{!}{2in}{\includegraphics{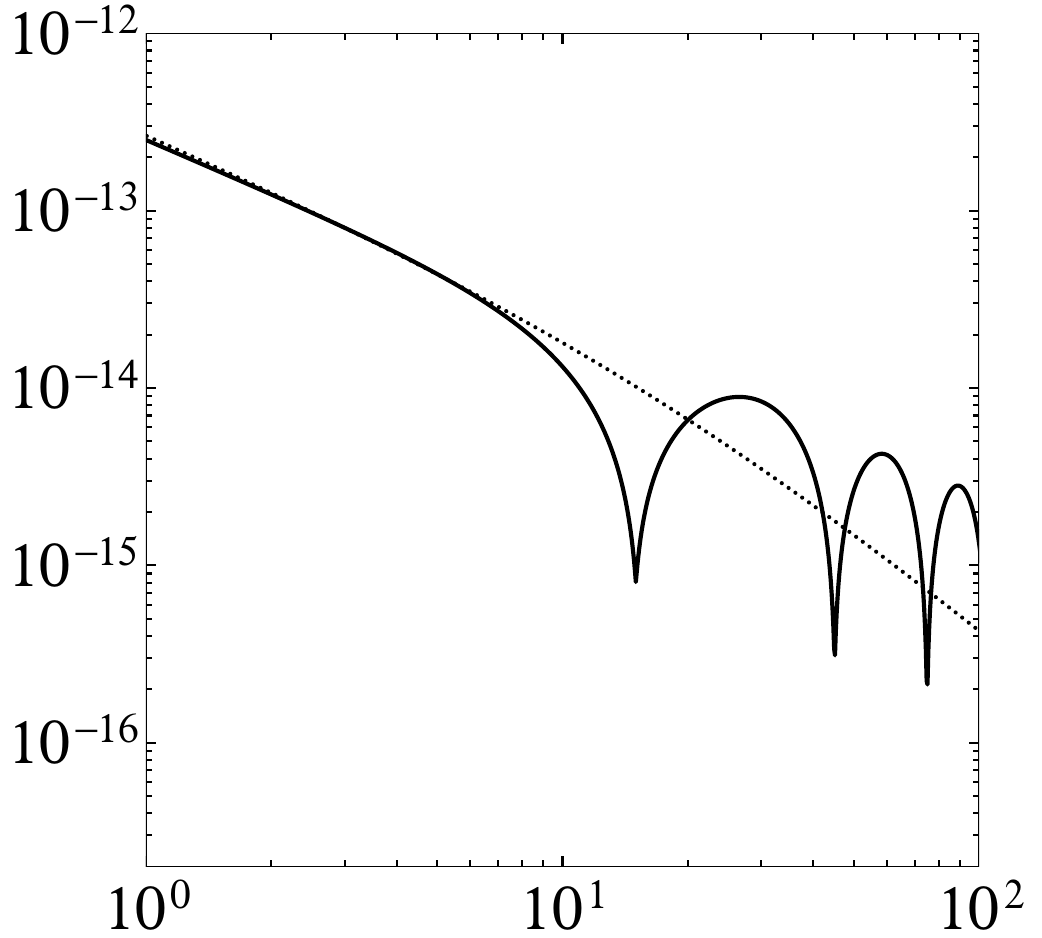}} \\
      & $f$ (Hz)
    \end{tabular}
  \end{center}
  \vspace{-2ex}
  \caption{\label{fig:fiorucci-xfer} Transfer function relating Newtonian acceleration noise to infrasound pressure noise, where the test mass is shielded from infrasound by a hemisphere of radius $R=6\,\mathrm{m}$ (equivalent effective underground depth $D=3\,\mathrm{m}$).  Solid curve: result from Figure~9 of~\cite{fiorucci:2018}.  Dotted line: result from Eq.~\ref{eq:sh-semiplane}, converted to equivalent units.  The results are broadly consistent, with differences due to the differing treatment of the wave's interaction with the structure.}
\end{figure}

Their numerical treatment of shielding by buildings also compares well with our heuristic approach.  Figure~\ref{fig:fiorucci-xfer} shows their ``transfer function'' (Figure~9 of~\cite{fiorucci:2018}) compared to our Eq.~\ref{eq:sh-semiplane}, adjusted 
account for different definitions (multiplied and divided by 2, then multiplied by $L\omega^2$ to get gravitational acceleration noise).  Per the discussion in section~\ref{ss:surface-structures}, we take $D=\frac{1}{2}R=3\,\mathrm{m}$ to compare with their hemispherical building with $R=6\,\mathrm{m}$.  Our agreement with their results is exact for low frequencies, and rolls off at a similar cutoff frequency.  The two results differ at high frequencies due to limitations in both approaches in modeling the interaction of the wave with the structure.  Our approach treats the wave as diffracting smoothly around the structure, producing a similarly smooth cutoff in the spectrum.  Their approach treats the wave as vanishing at one side and re-emerging on the other, producing resonant peaks when the structure is an exact multiple of the wavelength.  Neither approach is completely realistic in dealing with diffraction, reflection, and absorption (noise shadows).

We also note that Figure~9 of~\cite{fiorucci:2018} includes a transfer function for the \emph{interior} noise within the facility, which we do not consider in this paper.  We are primarily concerned with \emph{immitigable} noise from the environment, treating interior noise as, in principle, within the experimenter's control (although in practice reducing such noise may be a non-trivial task).

\section{Thermal Fluctuations}
\label{s:thermal}

Temperature fluctuations can cause density perturbations on the scales
of interest that are many orders of magnitude greater than pressure
fluctuations: typically on the order of $\sim10^{-3}$ fractional
changes (rather than $\sim10^{-7}$ for pressure waves).  However,
temperature fluctuations propagate past the instrument largely through
advection, at speeds of a few metres per second, rather than as waves
traveling at sonic speeds $\sim300\,\mathrm{m}/\mathrm{s}$.
In~\cite{creighton:2008} the effect of thermal perturbations was found to be significant only
if the airflow had turbulence that generated velocity fluctuations on
timescales corresponding to the frequencies of interest.

As with infrasound, we will start by (re)deriving the general scaling
law, before quoting the more precise result
from~\cite{creighton:2008}.  We then extend the analysis to include
parameters relevant to advanced gravitational-wave detectors, and show
how the noise is likely to behave in typical and worst-case scenarios.

Consider a packet of air of size $l$, with temperature perturbation $\delta T$ giving rise to density perturbation $\delta\rho=-\rho\delta T/T$.  The gravitational perturbation of one such packet is $g(t)=-G\rho l^3(\delta T/T)x(t)/r^3(t)$ where $r$ is the distance to the packet and $x$ is the component along the detector arm.  Now in general the noise power spectral density of a process $a(t)$ repeating at
intervals $\Delta t$ is $S_a(|f|)=(2/\Delta t)|\tilde{a}(f)|^2$.  So for a single wind streamline $i$, packets of air pass at intervals $l/v$ where $v$ is the wind speed, giving a noise contribution:
$$
S_{gi}(|f|) \;\;=\;\; \frac{2v}{l}\,|\tilde{g}(f)|^2 \;\;=\;\;
\frac{2v}{l}\,G^2\rho^2l^6\left(\frac{\delta T}{T}\right)^{\!2}\,
\left|\tilde{G}_S(f)\right|^2
$$
where $\tilde{G}_S(f)$ is the Fourier transform of
$G_S(t)=x(t)/r^3(t)$ along a streamline $S$.  Assuming the temperature
perturbations in streamlines more than $\sim l$ apart are incoherent,
then they will simply add in power, giving a total acceleration noise
spectrum:
$$
S_{g}(|f|) \;\;\sim\;\; \frac{2v}{l}\,G^2\rho^2l^6
\left(\frac{\delta T}{T}\right)^{\!2}
\int_{\{S\}} \left|\tilde{G}_S(f)\right|^2\,\frac{dA}{l^2}
$$
where $\{S\}$ is a plane intersecting all streamlines.  On small
scales, turbulent mixing gives rise to temperature fluctuations
between points that increase as the distance between those points
raised to some power $p$, typically $2/3$: $\delta T^2\sim c_T^2l^p$
(called the \textit{temperature structure function}).  So:
$$
S_{g}(|f|) \;\;\sim\;\; 2vG^2\rho^2\,\frac{c_T^2}{T^2}\,l^{p+3}
\int_S \left|\tilde{G}(f)\right|^2\,dA
$$
This formula still treats the air packet lengthscale $l$ as a fixed
parameter, and suggests that packets with larger $l$ contribute more
strongly to the gravitational noise.  However, for each frequency
there is a largest $l_\mathrm{max}\sim v/\omega$ that will contribute
to noise at that frequency.  This is because larger packets will
violate the point-mass approximation that went into the computation of
the gravitational field: $G_S(t)$ will be smoothed out on timescales
shorter than $l/v$, and thus $\tilde{G}_S(f)$ will be cut off at
frequencies higher than $v/l$.  To order of magnitude one can estimate
the noise by substituting $l$ with this maximum $v/\omega$ to obtain:
$$
\sqrt{S_g} \;\;\sim\;\; G\rho\,\sqrt{\frac{c_T^2}{T^2}\,
  \frac{v^{p+4}}{\omega^{p+3}}\int_S \left|\tilde{G}_S(f)\right|^2dA}
$$
Taking $\tilde{h}=\tilde{g}/L\omega^2$ as before, the strain noise
spectrum is:
$$
\sqrt{S_h} \;\;\sim\;\; \frac{G\rho}{L}\,\sqrt{\frac{c_T^2}{T^2}\,
  \frac{v^{p+4}}{\omega^{p+7}}\int_S \left|\tilde{G}_S(f)\right|^2dA}
$$
Applying a similar argument to a gravity gradiometer gives:
$$
\sqrt{S_h} \;\;\sim\;\; G\rho\,\sqrt{\frac{c_T^2}{T^2}\,
  \frac{v^{p+4}}{\omega^{p+7}}\int_S \left|\tilde{G}_S'(f)\right|^2dA}
$$
where $\tilde{G}_S'(f)$ is the Fourier transform of
$G_S'(t)=(r^2-3x^2)/r^5$.

\begin{figure}
  \begin{center}
    \begin{tabular}{cc}
      \rotatebox{90}{\makebox[2in][c]{$\sin(p\pi/2)\,\Gamma(p+1)$}} &
      \resizebox{!}{2in}{\includegraphics{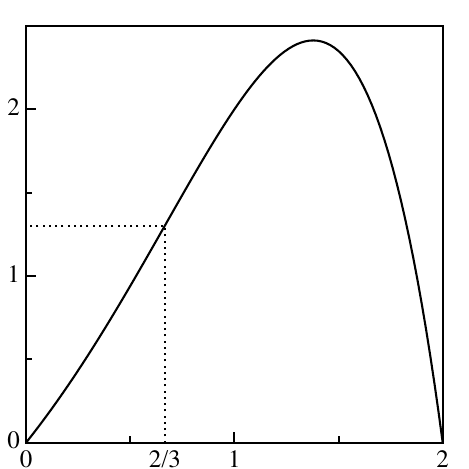}} \\
      & $p$
    \end{tabular}
  \end{center}
  \vspace{-2ex}
  \caption{\label{fig:ap} Plot of $A(p)=\sin(p\pi/2)\,\Gamma(p+1)$,
    the $p$-dependent scale factor from Eq.~(\ref{eq:sh-general}),
    with the canonical value $p=2/3$ indicated.  To a reasonable
    approximation, this factors can be treated as $\sim1$ over typical
    ranges of $p$.}
\end{figure}

In~\cite{creighton:2008} the more precise derivation gives:

\begin{equation}
  \label{eq:sh-general}
  S_h \;\;=\;\; \frac{2\pi^2G^2\rho^2}{L^2}\,\frac{c_T^2}{T^2}\,\omega^{-(p+7)}
  A(p)\,\sum_k\int_{\{S\}}\tilde{F}_{S,k}(f)^*\tilde{G}_{S,k}(f)\,v_0\,dA
\end{equation}

\noindent where $A(p)=\sin(p\pi/2)\Gamma(p+1)$ represents the weak
(non-power-law) dependence on $p$, the sum is over individual test
masses $k$, $\tilde{F}_{S,k}(f)$ is the Fourier transform of
$F_{S,k}(t)=G_{S,k}(t)v(t)^{p+3}$, where $v(t)$ is allowed to vary
along a streamline, and $v_0$ is the value of $v(t)$ as it crosses the
plane of integration $\{S\}$.  Note that the functional form of $A(p)$
arises from certain formal assumptions made in the derivation; while
these assumptions may not be exactly met in practice, one would expect
that a power law dependence in $\omega$ with leading factor
$A(p)\sim1$ should hold approximately true in general.  The formal
function is plotted in figure~\ref{fig:ap}, also indicating the
canonical value $A(p)\approx1.3$ for $p=2/3$.

In~\cite{creighton:2008} the functions $\tilde{F}_{S,k}(f)$ and
$\tilde{G}_{S,k}(f)$ were then evaluated over various specific, but
ultimately arbitrary, geometries.  It should also be noted that these
scenarios made no connection between the motion $\bb{r}(t)$ along a
streamline, and the turbulent convection that created the power-law
spectrum of temperature perturbations in the first place.  Instead,
the temperature perturbations were treated as if they were
pre-established according to the specified power law and then
``frozen'' into the advected airflow.  It is worth reiterating, as
in~\cite{creighton:2008}, the need for fluid dynamic simulations of or
measurements from actual or proposed detector sites -- something that
is beyond the scope of this paper.  Instead, the following section
will apply the above approximations to get typical and
worst-case noise estimates with analytic parameters that can be tuned
to specific detector design proposals.

\subsection{Smooth versus turbulent airflow}
\label{ss:smooth-vs-turbulent-airflow}

For smooth uniform airflow, one can write $x(t)=x_0+v_xt$,
$r(t)=\sqrt{r_0^2+v^2t^2}$, $v(t)=v$, where we arbitrarily choose
$t=0$ to be the time the of closest approach between the air element
and the test mass, $r_0$ and $x_0$ are the distance and the
$x$-position at that time, $v$ is the speed of the airflow, and $v_x$
is the velocity component along the $x$-axis.  The Fourier transform
is straightforward:

\begin{equation}
  \label{eq:g-smooth}
  \tilde{G}_S(f) \;\;=\;\; \frac{2\omega}{v^2}\left[
    \frac{x_0}{r_0}\,K_1\!\left(\frac{\omega r_0}{v}\right) -
    i\frac{v_x}{v}\,K_0\!\left(\frac{\omega r_0}{v}\right)\right] \;,
  \qquad \omega=2\pi f
\end{equation}

\noindent where $K_0$ and $K_1$ are modified Bessel functions of the
second kind of order 0 and 1.  For the frequencies of interest for
advanced gravitational wave detectors, $\omega\sim10$,
$r_0\sim10\,\mathrm{m}$, $v\lessim10\,\mathrm{m}/\mathrm{s}$, so the
arguments of these Bessel functions are moderately large, and one can
approximate $K_0(z)\sim K_1(z)\sim\sqrt{\pi/2z\,}e^{-z}$:
$$
|\tilde{G}_S(f)|^2 \;\;\sim\;\; \frac{2\pi\omega}{r_0 v^3}
\left(\frac{x_0^2}{r_0^2}+\frac{v_x^2}{v^2}\right)e^{-2\omega r_0/v}
$$
which is exponentially cut off at the frequencies of interest.

However, as shown in~\cite{creighton:2008}, much stronger signals at a
given frequency can arise if the airflow has circulation patterns or
turbulence with periodicities around the frequencies of interest.  In
that paper, a specific but somewhat arbitrary circulation pattern was
adopted as illustration.  Here we present a more general argument.

First, consider a worst-case scenario where all flowlines have
circulation with radius $R$ and angular frequency $\Omega\sim v/R$ (in
the case of equality this corresponds to strictly cycloidal motion).
To a reasonable approximation, we simply add a component $R_x\cos\Omega
t$ to $x(t)$, leaving $r(t)$ and $v(t)$ unchanged.  Here $R_x=R\cos\theta$ where $\theta$ is the angle of the airflow to the interferometer measurement axis.  Unlike the case of infrasound we do \emph{not} average over this angle, as it is not a stochastic property.  By the convolution
theorem this adds a component to $\tilde{G}_S(f)$ similar to the $K_1$
term but offset in frequency (and reduced by a factor 1/2):
$$
\tilde{G}_S(f) \;\;\approx\;\; \frac{|\omega-\Omega|}{v^2}\,
\frac{R_x}{r_0}\,K_1\!\left(\frac{|\omega-\Omega|r_0}{v}\right)
$$
This function is peaked in a band $\Delta\omega\approx v/r_0$ about
$\omega=\Omega$, falling off exponentially outside of this band.  Note
that we are implicitly ignoring a symmetric peak about $\omega=-\Omega$,
as well as the previous terms peaked about $\omega=0$, on the
assumption that the cutoff will make these other terms negligible.  We
also note that $K_1(z)\approx1/z$ for small $z$, so the \emph{peak}
value of $\tilde{G}_S(f)$ is $\approx R_x/r_0^2 v$.

To integrate this over all streamlines, we adopt the same geometry used
for the infrasound derivation, placing the test mass a distance $D$
below ground and integrating over the airmass above ground; it is
understood that other detector geometries will have a similar form
(though with different prefactors), where $D$ represents the minimum
distance between the test mass and the airflow.  As before we also cut
off the integration at the boundary layer height $H$, which may be
significant for deeply-buried detectors.  Then:
$$
\begin{array}{rcl}
  \displaystyle\int_{\{S\}}\tilde{F}_{S,k}(f)^*\tilde{G}_{S,k}(f)\,v_0\,dA
  &\approx& \displaystyle v^{p+4}\int_{\{S\}}\left|\tilde{G}_{S,k}(f)\right|^2dA
  \;\;\approx\;\; v^{p+4}\int_{\{S\}}\frac{R_x^2}{r_0^4\,v^2}\,dA \\[2.5ex]
  &\approx& \displaystyle v^{p+4}\int_{y=-\infty}^\infty\int_{z=D}^{D+H}
  \frac{R_x^2}{(y^2+z^2)^2v^2}\,dy\,dz \\[3ex]
  &\approx& \displaystyle \frac{\pi}{4}\,R_x^2\,v^{p+2}
  \left(\frac{1}{D^2}-\frac{1}{(D+H)^2}\right)
\end{array}
$$
Note that the final parenthetical term goes as $1/D^2$ for $D\ll H$
and as $2H/D^3$ for $D\gg H$, so as before we write it approximately as
a cut-off power law $D^{-2}(1+D/2H)^{-1}$, a computationally safer
expression that more readily shows these limits.  Assuming the four
test masses of an interferometer contribute equally, the noise in $h$
is then:
$$
S_h \;\;\approx\;\; \frac{2\pi^2G^2\rho^2}{L^2}\,\frac{c_T^2}{T^2}\,
\frac{v^{p+2}}{\omega^{p+7}}A(p)\,\frac{R_x^2}{D^2}
\left(1+\frac{D}{2H}\right)^{-1}
$$
Note that this is the \emph{peak} noise at the frequency
$\omega=\Omega\sim v/R$.  Substituting this expression in for $R$
gives the \emph{locus} of how high the peak noise might reach at any
given frequency:

\begin{equation}
  \label{eq:sh-peak}
  S_h \;\;\lessim\;\; \frac{2\pi^2\cos^2\theta G^2\rho^2}{D^2L^2}\,\frac{c_T^2}{T^2}\,
  \frac{v^{p+4}}{\omega^{p+9}}A(p)\left(1+\frac{D}{2H}\right)^{-1}
\end{equation}

\noindent where the maximum width of the peak (based on packets that
pass closest to the test mass) is $\Delta\omega\sim v/D$.

More typically one would expect turbulence on a \emph{range} of
frequency scales.  By loose analogy with the infrasound case, we use a
``quality factor'' $Q$ to parameterize coherence, not of the
underlying temperature perturbations, but of the cyclic motions in the
airflow.  That is, we assume the noise power is spread over a
bandwidth $\Delta\omega\sim\omega/Q$, and the power level drops by a
factor $\sim Qv/\omega D$, where $Q\sim1$ represents the more
``typical'' case of a broad spectrum of turbulence, giving:

\begin{equation}
  \label{eq:sh-broad}
  S_h \;\;\sim\;\; \frac{2\pi^2\cos^2\theta G^2\rho^2}{D^3L^2}\,\frac{c_T^2}{T^2}\,
  \frac{Qv^{p+5}}{\omega^{p+10}}A(p)\left(1+\frac{D}{2H}\right)^{-1}
\end{equation}

\noindent Equations~(\ref{eq:sh-peak}) and~(\ref{eq:sh-broad}), for
$Q\sim1$, merge around $\omega\sim v/D$, which is also where the
smooth-airflow formula in equation~(\ref{eq:g-smooth}) begins to
dominate.  When $\omega D/v\lessim1$, equations~(\ref{eq:g-smooth}) is
dominated by the term $K_1(z)\sim1/z$ for small $z$: after integrating
over a half-plane starting a distance $D$ from the test mass this
results in a \emph{logarithmic} frequency dependence:
$$
S_h \;\;\sim\;\; \frac{4\pi^3\cos^2\theta G^2\rho^2}{L^2}\,\frac{c_T^2}{T^2}\,
\frac{v^{p+2}}{\omega^{p+7}}\ln\left(\frac{v}{\omega D}\right)A(p)
$$
One would expect this last expression also to have an additional
cutoff when $D\gtrsim H$.  However, since the expression is for
$D\lessim v/\omega\sim l$, this would also imply $l\gtrsim H$;
i.e.\ it would assume that the noise was dominated by the contribution
of thermal pockets larger than the convective boundary layer, which
stretches the model's assumptions.  Instead of extrapolating the
logarithmic form of this expression, we instead apply it as an
approximate cutoff to equations~(\ref{eq:sh-peak})
and~(\ref{eq:sh-broad}) using the Fermi problem rule that ``the
logarithm of any large number is 10'', to obtain:

\begin{eqnarray}
  S_h &\sim& \frac{2\pi^2\cos^2\theta G^2\rho^2}{L^2}\,\frac{c_T^2}{T^2}\,
  A(p)\,\left(\frac{v}{\omega}\right)^{p+2}\omega^{-5}
  \left(1+\frac{D}{2H}\right)^{-1} \\
  && \qquad\times\left(\frac{1}{3}+\frac{\omega D}{v}\right)^{-2}
  \left(1+\frac{\omega D}{Qv}\right)^{-1} \nonumber
\end{eqnarray}

\noindent In this formulation, setting $Q\gg1$ recovers the behaviour
of equation~(\ref{eq:sh-peak}): the maximum noise assuming a turbulent
flow with spectrum peaked strongly at a given frequency of interest.
$Q\sim1$ corresponds to a flow with strong turbulence at the frequency
of interest, but spread over a broad band $\Delta\omega\sim\omega$.
Smaller values of $Q$ can represent the case where the turbulent
spectrum is not peaked at all, but is spread over several orders of
magnitude in frequency (including the frequency of interest): for
instance, $Q\sim0.1$ might represent a broad turbulent spectrum
spanning 10 $e$-foldings in frequency.

For a gradiometric gravitational-wave sensor, one can substitute
$G_S'(t)=R_x^2\cos^2(\Omega t)r(t)^{-5}$ in place of
$G_S(t)/L=(R_x/L)\cos(\Omega t)r(t)^{-3}$.  To order of magnitude, an
interferometer and a gradiometer will have similar noise when these
two functions are comparable, which occurs when $L\sim r^2/R_x\sim
D^2\omega/v$.  Noting that $\cos^2\Omega t=(1+\cos2\Omega t)/2$. one
can see that $\tilde{G}_S'(f)$ will be peaked at \emph{twice} the
circulation frequency:
$$
\tilde{G}_S'(f) \;\;\approx\;\; \frac{(\omega-2\Omega)^2}{6v^3}\,
\frac{R_x^2}{r_0^2}\,K_2\!\left(\frac{(\omega-2\Omega)r_0}{v}\right)
$$
Now $K_2(z)$ has the same exponential fall-off as $K_1(z)$, but for
\emph{small} $z$ one has $K_2(z)\approx2/z^2$, and thus the
\emph{peak} value of $\tilde{G}_S'(f)$ is $R_x^2/3vr_0^4$ at
$\omega=2\pi f=2\Omega$, with a bandwidth $\Delta\omega\approx v/D$.
Integrating over the half-plane a distance $D$ above the test mass
gives:
$$
\int_{\{S\}}\tilde{F}_{S,k}'(f)^*\tilde{G}_{S,k}'(f)\,v_0\,dA \;\;\approx\;\;
\frac{5\pi}{864}\,R_x^4\,v^{p+2}\left(\frac{1}{D^6}-\frac{1}{(D+H)^6}\right)
$$
With the approximation $D^{-6}-(D+H)^{-6}\approx D^{-6}(1+D/6H)^{-1}$
and the substitution $\omega=2\Omega\sim2v/R$ this gives an overall
strain noise spectrum of:

\begin{eqnarray}
  \label{eq:sh-grad-peak}
  S_h &\lessim& \frac{5\pi^2\cos^4\theta G^2\rho^2}{27D^6}\,\frac{c_T^2}{T^2}\,
  \frac{v^{p+6}}{\omega^{p+11}}A(p)
  \left(1+\frac{D}{6H}\right)^{-1} \quad \Delta\omega\sim v/D\\
  \label{eq:sh-grad-broad}
  &\sim& \frac{5\pi^2\cos^4\theta G^2\rho^2}{27D^7}\,\frac{c_T^2}{T^2}\,
  \frac{Qv^{p+7}}{\omega^{p+12}}A(p)
  \left(1+\frac{D}{6H}\right)^{-1} \quad \Delta\omega\sim\omega/Q
\end{eqnarray}

\noindent These follow equations~(\ref{eq:sh-peak})
and~(\ref{eq:sh-broad}) but with a relative scaling of
$5L^2v^2/54D^4\omega^2$.  They also have a slightly different cutoff
as $D\gtrsim H$, and also as $\omega\lessim v/D$; however, these
cutoffs are only approximate anyway.  Therefore one can combine all
four equations into a single approximation that applies whether the
arm length is longer or shorter than the characteristic scale
$D(D\omega/v)$ over which the gravitational field is varying:

\begin{eqnarray}
  \label{eq:sh-thermal}
  S_h &\sim& \frac{2\pi^2\cos^2\theta G^2\rho^2}{D^4}\,\frac{c_T^2}{T^2}\,
  A(p)\,\left(\frac{v}{\omega}\right)^{p+4}\omega^{-5}
  \left(\frac{\sqrt{54/5}}{\cos\theta}+\frac{Lv}{D^2\omega}\right)^{-2} \\
  && \qquad\times\left(1+\frac{D}{2H}\right)^{-1}
  \left(\frac{1}{3}+\frac{\omega D}{v}\right)^{-2}
  \left(1+\frac{\omega D}{Qv}\right)^{-1} \nonumber
\end{eqnarray}

\subsection{Temperature Measurements}
\label{ss:temperature-measurements}

In~\cite{coulman:1985}, typical daytime peak temperature perturbations
have $c_T^2\approx0.2\,\mathrm{K}^2\mathrm{m}^{-2/3}$ with $p=2/3$.
In practice, however, one will want to measure the actual temperature
structure function at the site.  One method is simply to use a
rapid-response thermometer to measure temperature perturbations as
they are advected past a point in space, along with the speed of the
airflow $v$.  From the derivation in~\cite{creighton:2008}, the
frequency spectrum of temperature perturbations is then:
\begin{eqnarray}
  S_T &=& c_T^2v^p\omega^{-(p+1)}\sin(p\pi/2)\Gamma(p+1) \nonumber\\
  \label{eq:st}
  &=& c_T^2\,\left(\frac{v}{\omega}\right)^p\omega^{-1}\,(p+1)\,A(p)
\end{eqnarray}

\noindent For the typical parameters from~\cite{coulman:1985} this
gives
$\sqrt{S_T}\approx(0.14\,\mathrm{K}/\sqrt{\mathrm{Hz}})(v/(\mathrm{m}/\mathrm{s}))^{1/3}(f/\mathrm{Hz})^{-5/6}$.

Assuming simultaneous measurement of $\sqrt{S_T}$ and $\sqrt{S_h}$,
i.e.\ using the same $v$ in both equations~(\ref{eq:sh-thermal})
and~(\ref{eq:st}), the estimated gravitational noise spectral density
(for turbulent advection) can be written directly in terms of the
measured temperature spectral density:
\begin{eqnarray}
  \sqrt{S_h} &\approx& \sqrt{\mbox{$\frac{5}{3}$}(p+1)}\,\pi\cos\theta G\rho\,
  \frac{\sqrt{S_T}}{T}\,\frac{v^2}{D^2\omega^4}
  \left(1+\sqrt{\frac{5}{54}}\,\frac{vL\cos\theta}{D^2\omega}\right)^{-1} \nonumber\\
  && \qquad\times\left(1+\frac{D}{2H}\right)^{-1/2}
  \left(1+\frac{\omega D}{3v}\right)^{-1}
  \left(1+\frac{\omega D}{Qv}\right)^{-1/2} \nonumber
\end{eqnarray}

\noindent This expression still has a formal dependence on the power
law index $p$, which presumes that $S_T$ is an exact power law.
However, to a reasonable approximation we can simply use the canonical
value $p=2/3$ even when $S_T$ may not have an ideal power law form.
This choice of $p$ (coincidentally) simplifies the initial prefactor;
we also approximate $\sqrt{5/54}\approx1/3$ to obtain:
\begin{eqnarray}
  \label{eq:sh-t-final}
  \sqrt{S_h} &\approx& \mbox{$\frac{5}{3}$}\,\pi\cos\theta G\rho\,
  \frac{\sqrt{S_T}}{T}\,\frac{v^2}{D^2\omega^4}
  \left(1+\frac{vL\cos\theta}{3D^2\omega}\right)^{-1} \\
  && \qquad\times\left(1+\frac{D}{2H}\right)^{-1/2}
  \left(1+\frac{\omega D}{3v}\right)^{-1}
  \left(1+\frac{\omega D}{Qv}\right)^{-1/2} \nonumber
\end{eqnarray}

We take note of some key differences between this expression and the corresponding Eq.~(\ref{eq:sh-incoherent}) for infrasound.  Thermal fluctuations have a steeper power law in $\omega$, especially considering that the various factors $\omega D/v$ are much larger than the corresponding factors $\omega D/v_s$ for infrasound.  However, thermal fluctuations have a much stronger driving factor $\sqrt{S_T}/T\gg\sqrt{S_P}/P$.  Furthermore, we note the absence of exponential cutoffs in the thermal spectrum: the case we are concerned with here is one of spatially-coherent \emph{modulation} of an underlying stochastic fluctuation, which is both harder to shield against and harder to cancel than coherent perturbations.

\subsection{Comparison with Other Results}
\label{ss:comparison-with-other-results}

In~\cite{brundu:2022}, a numerical analysis is presented that extends the results from section~3.1 of~\cite{creighton:2008}: they consider the case of \emph{uniform} airflow, but relax the assumption that temperature perturbations are ``frozen in'' to the moving airmass, and instead introduce a more realistic model for thermal convection.  They find, as above, that the exponential behaviour $\sim e^{-\omega D/v}$ gives way to a power law at high frequencies, which can be understood as being due to the fact that convective turbulence introduces fluctuations on timescales shorter than the crossing time $D/v$.  However, they find that this transition typically occurs well below the sensitivities of 3rd Generation detectors.

Our analysis here instead extends the results of section~3.3 of~\cite{creighton:2008}, where temperature perturbations are treated as frozen and turbulence is directly imposed on the airflow streamlines.  This turbulence \emph{need not be due to thermal convection}, but, more seriously, could arise from mechanical interaction with the ground or surface structures.  A predictive numerical model including these effects would depend sensitively on site-specific and environmental conditions, so our model instead introduces (via $Q$) a range of typical to worst-case scenarios.  The result, similar to~\cite{brundu:2022}, is a transition from exponential to power law behaviour, but potentially at lower frequencies and higher amplitudes.  Large-scale mechanical turbulence could therefore have a measurable impact on 3rd Generation detectors, which is discussed further in the next section.

\section{Standard Cases}
\label{s:cases}

Now we will consider some typical design parameters for advanced
interferometers and see how much one might expect atmospheric
gravitational noise to contribute to the instrument.  In some cases,
even the precise location and geography of the site has not yet been
determined; therefore we will assume common values for the underlying
pressure and temperature fluctuations.  In keeping with pessimistic
``worst case'' assumptions, we assume an underlying infrasound noise
spectrum from the $95^\mathrm{th}$ percentile noon-time spectra
in~\cite{woodward:2005} (assuming it is entirely coherent):
$$
\sqrt{S_P} \;\;=\;\; \left(0.1\,\mathrm{Pa}/\sqrt{\mathrm{Hz}}\right)
\left(\frac{f}{\mathrm{Hz}}\right)^{-1}
$$
For temperature perturbations, we assume typical daytime peak values
from~\cite{coulman:1985}) and moderately high winds
$v=10\,\mathrm{m}/\mathrm{s}$, giving
$$
\sqrt{S_T} \;\;=\;\; \left(0.3\,\mathrm{K}/\sqrt{\mathrm{Hz}}\right)
\left(\frac{f}{\mathrm{Hz}}\right)^{-5/6}
$$
We assume sea level values of $\rho=1.2\,\mathrm{kg}/\mathrm{m}^3$,
$v_s=340\,\mathrm{m}/\mathrm{s}$, and $P=1.0\times10^{5}\mathrm{Pa}$,
and assume an ambient temperature of $T=300\,\mathrm{K}$ and
convective boundary layer height $H=1\,\mathrm{km}$.  Pessimistically we take $\cos\theta=1$.  In terms of the
remaining design parameters $D$ and $L$ the parameterized spectra are:
\begin{eqnarray}
  \makebox[0pt][r]{Sonic:\quad}
  \sqrt{S_h} &=& \left(6.4\times10^{-18}/\sqrt{\mathrm{Hz}}\right)\!
  \left(\frac{f}{\mathrm{Hz}}\right)^{\!\!-3}\!
    \left(1+\frac{fL}{108\,\mathrm{m}/\mathrm{s}}\right)^{-1}\\
    && \times\left(1+\frac{0.2}{Q^5}\left(\frac{fD}{54\,\mathrm{m}/\mathrm{s}}\right)^3\left(1+\frac{D}{3000\,\mathrm{m}}\right)\right)^{-1/2}
  \nonumber\\
      && \times\left(e^{-fD/(54\,\mathrm{m}/\mathrm{s})}+e^{-3.1Q}\right)
  \nonumber\\[2ex]
  \makebox[0pt][r]{Thermal:\quad}
  \sqrt{S_h} &=& \left(2.7\times10^{-14}/\sqrt{\mathrm{Hz}}\right)\!
  \left(\frac{D}{\mathrm{m}}\right)^{\!\!-2}\!
  \left(\frac{f}{\mathrm{Hz}}\right)^{\!\!-29/6}\!
  \left(1+\frac{0.5\mathrm{m}/\mathrm{s}}{D^2f/L}\right)^{\!\!-1} \\
  && \times\left(1+\frac{D}{2000\,\mathrm{m}}\right)^{\!\!-1/2}\!
  \left(1+\frac{fD}{5\,\mathrm{m}/\mathrm{s}}\right)^{\!\!-1}\!
  \left(1+\frac{fD/Q}{1.6\,\mathrm{m}/\mathrm{s}}\right)^{\!\!-1/2} \nonumber
\end{eqnarray}

\noindent For infrasound, we plot curves for moderately-coherent ($Q=1$) and highly-coherent ($Q\rightarrow\infty$) pressure fields.  We note that highly-coherent infrasound will generally produce \emph{lower} Newtonian noise in any underground detector, but in all the cases considered below, the effect of incoherence is below the threshold of sensitivity.  For thermal
noise, we similarly plot a range between two cases: an underlying spectrum
assuming broadband turbulence over many $e$-foldings of frequency
($Q=0.1$), and the locus of maxima where the noise might reach if the
turbulent spectrum is peaked in a narrow band ($Q\rightarrow\infty$).  In this case a coherent wind pattern produces \emph{higher} Newtonian noise, albeit in a narrow frequency band.

\subsection{Advanced LIGO}

\begin{figure}
  \begin{center}
    \begin{tabular*}{\hsize}{@{}c@{}c@{\extracolsep{\fill}}c@{\extracolsep{0pt}}c@{}}
      \rotatebox{90}{\makebox[2in][c]{$\sqrt{S_h}$ ($\mathrm{Hz}^{-1/2}$)}} &
      \resizebox{!}{2in}{\includegraphics{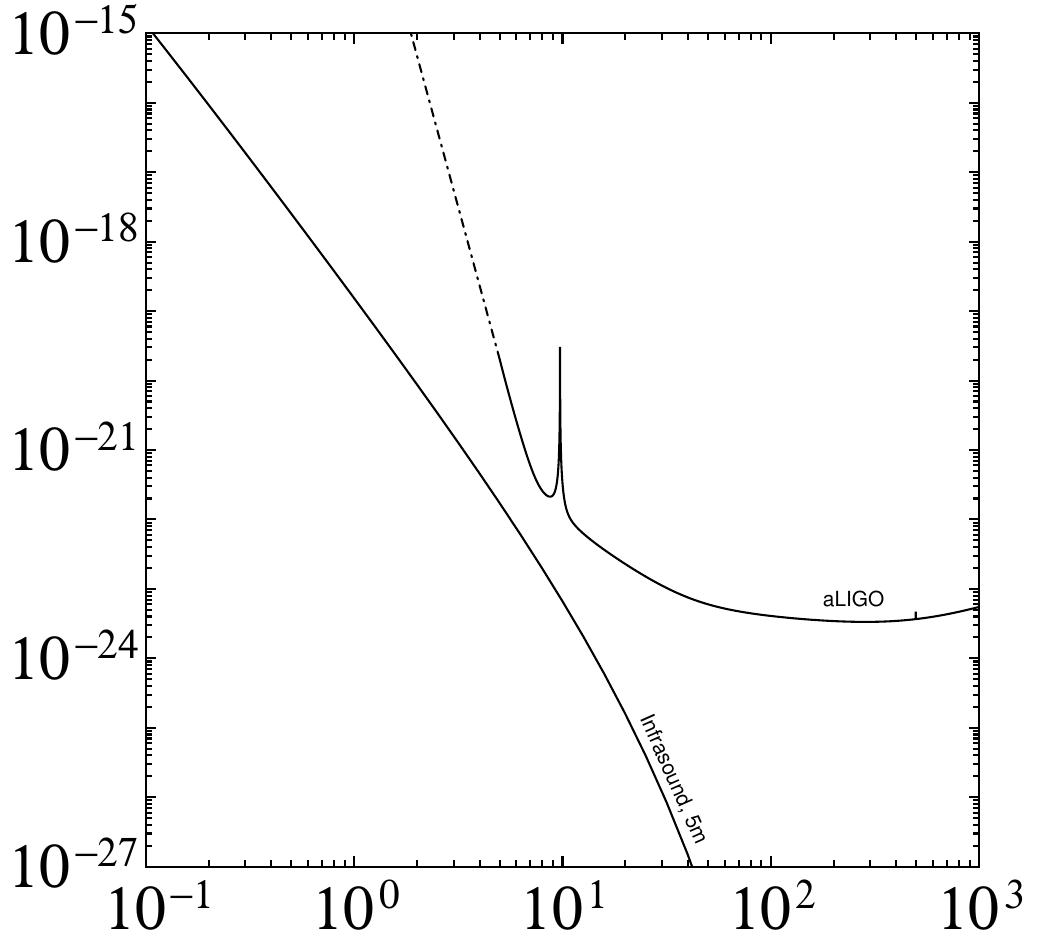}} &
      \rotatebox{90}{\makebox[2in][c]{$\sqrt{S_h}$ ($\mathrm{Hz}^{-1/2}$)}} &
      \resizebox{!}{2in}{\includegraphics{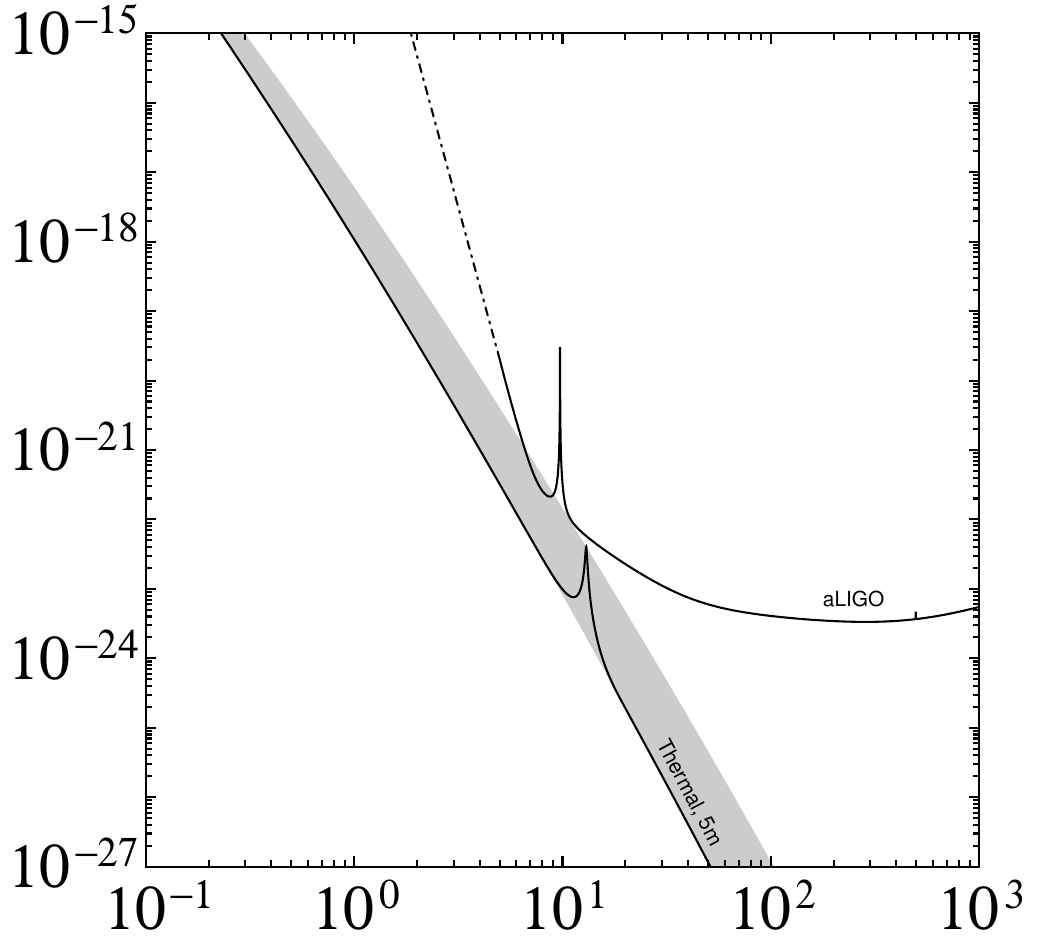}} \\
      (a) & $f$ (Hz) & (b) & $f$ (Hz)
    \end{tabular*}
  \end{center}
  \vspace{-2ex}
  \caption{\label{fig:aligo} Estimated Newtonian noise levels in
    Advanced LIGO from (a) infrasound and (b) thermal perturbations,
    assuming no perturbations within 5\,m of the test masses.  The
    line labeled aLIGO shows the detector design noise from~\cite{aligo-sensitivities} (dot-dashed lines are the authors' extrapolation).  For thermal
    Newtonian noise, the bottom of the shaded band corresponds to a
    broadband spectrum of turbulence, while the top of the shaded band
    is the locus of ``worst case'' noise levels for
    turbulence concentrated at a particular frequency: an example of
    such a spectrum is shown for an arbitrarily chosen peak frequency.}
\end{figure}

In Fig.~\ref{fig:aligo}, we plot the infrasound and thermal noise
spectra against the Advanced LIGO design sensitivity curve
from~\cite{aligo-sensitivities}, using the parameters $L=4\,\mathrm{km}$ and
$D=5\,\mathrm{m}$ (typical for the LIGO end stations).

From these plots, it appears that atmospheric Newtonian noise is not
likely to be a severe constraint on the target noise curve, though
thermal Newtonian noise may cut into the corner of the noise curve on
bad days.  It is also clear that these sources will constrain any
attempt to push to lower frequencies at the LIGO sites, unless
mechanisms can be found to mitigate them.

\subsection{Einstein Telescope}

\begin{figure}
  \begin{center}
    \begin{tabular*}{\hsize}{@{}c@{}c@{\extracolsep{\fill}}c@{\extracolsep{0pt}}c@{}}
      \rotatebox{90}{\makebox[2in][c]{$\sqrt{S_h}$ ($\mathrm{Hz}^{-1/2}$)}} &
      \resizebox{!}{2in}{\includegraphics{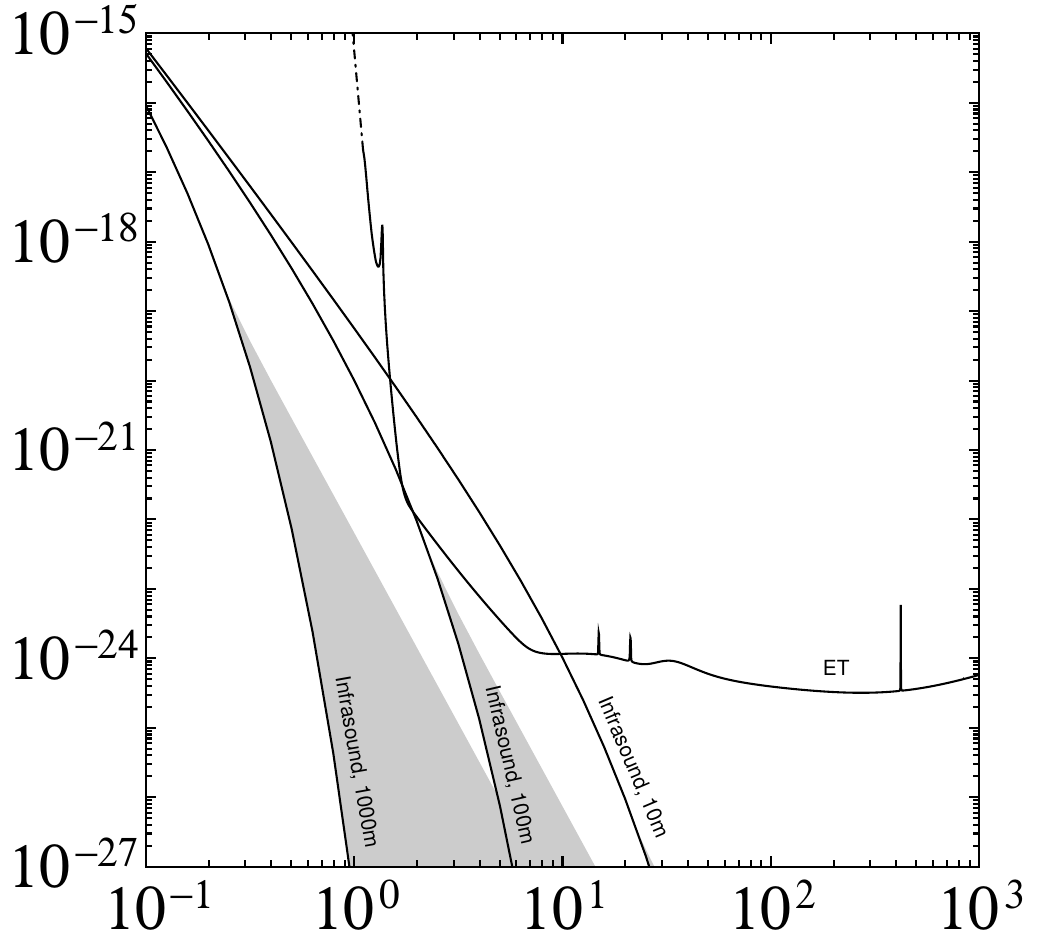}} &
      \rotatebox{90}{\makebox[2in][c]{$\sqrt{S_h}$ ($\mathrm{Hz}^{-1/2}$)}} &
      \resizebox{!}{2in}{\includegraphics{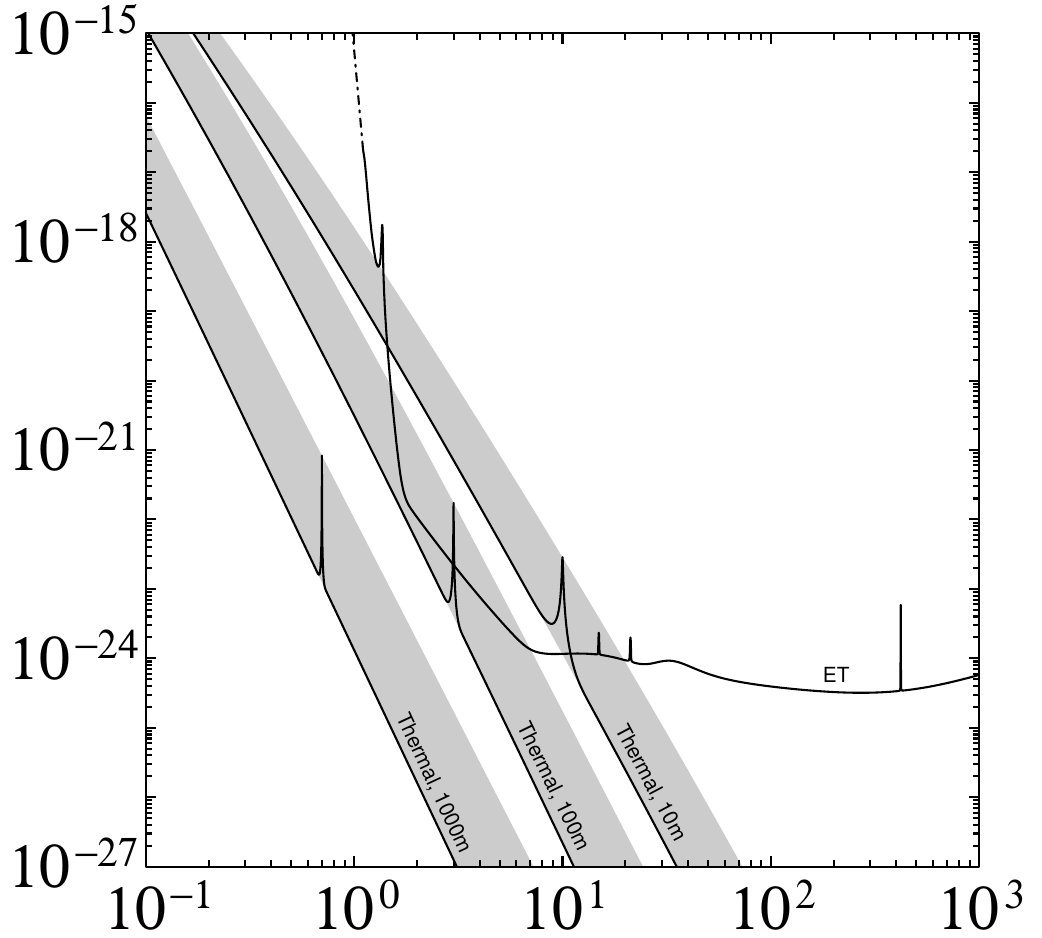}} \\
      (a) & $f$ (Hz) & (b) & $f$ (Hz)
    \end{tabular*}
  \end{center}
  \vspace{-2ex}
  \caption{\label{fig:et} Estimated Newtonian noise levels in the
    Einstein Telescope (ET) from (a) infrasound and (b) thermal
    perturbations.  In each case, three noise estimates are plotted,
    assuming no perturbations within 10\,m, 100\,m, or 1000\,m of the
    test masses, respectively.  The line labeled ET shows the detector
    design noise from~\cite{et-sensitivities} (dot-dashed lines are the authors' extrapolation). For infrasound, the solid curves correspond to highly coherent infrasound, while the tops of the shaded regions assume coherence lengths comparable to wavelength.  For thermal Newtonian noise, the bottoms of the shaded
    bands correspond to a broadband spectrum of turbulence, while the
    tops of the shaded bands are the loci of ``worst
    case'' noise levels for turbulence concentrated at a particular
    frequency: examples of such spectra are shown at arbitrarily
    chosen peak frequencies.}
\end{figure}

In Fig.~\ref{fig:et}, we plot the infrasound and thermal noise spectra
against the Einstein Telescope (ET) design sensitivity curve
from~\cite{et-sensitivities}, using the design length $L=10\,\mathrm{km}$.  The
proposed site of the observatory has not been determined, but it is
expected to be underground.  Noise curves are given assuming
perturbations are excluded within 10\,m, 100\,m, and 1\,km of the test
masses.

From figure~\ref{fig:et}(a), it seems likely that infrasonic Newtonian
noise may dominate the low-frequency corner of the target noise curve,
unless some mitigation mechanism is put in place.  From
figure~\ref{fig:et}(b), it is clear that the observatory should be
placed at least 100\,m underground in order to ensure that thermal
Newtonian noise does not dominate the noise budget.

\subsection{Cosmic Explorer}

\begin{figure}
  \begin{center}
    \begin{tabular*}{\hsize}{@{}c@{}c@{\extracolsep{\fill}}c@{\extracolsep{0pt}}c@{}}
      \rotatebox{90}{\makebox[2in][c]{$\sqrt{S_h}$ ($\mathrm{Hz}^{-1/2}$)}} &
      \resizebox{!}{2in}{\includegraphics{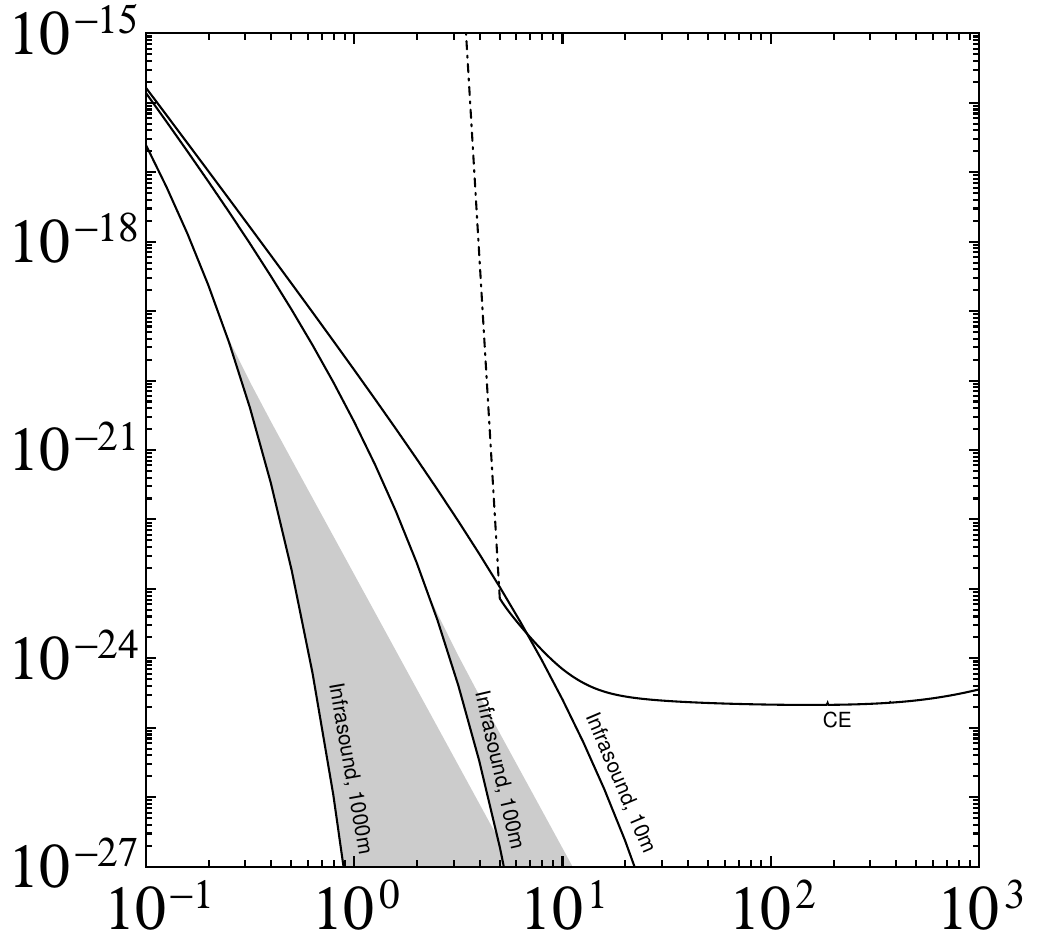}} &
      \rotatebox{90}{\makebox[2in][c]{$\sqrt{S_h}$ ($\mathrm{Hz}^{-1/2}$)}} &
      \resizebox{!}{2in}{\includegraphics{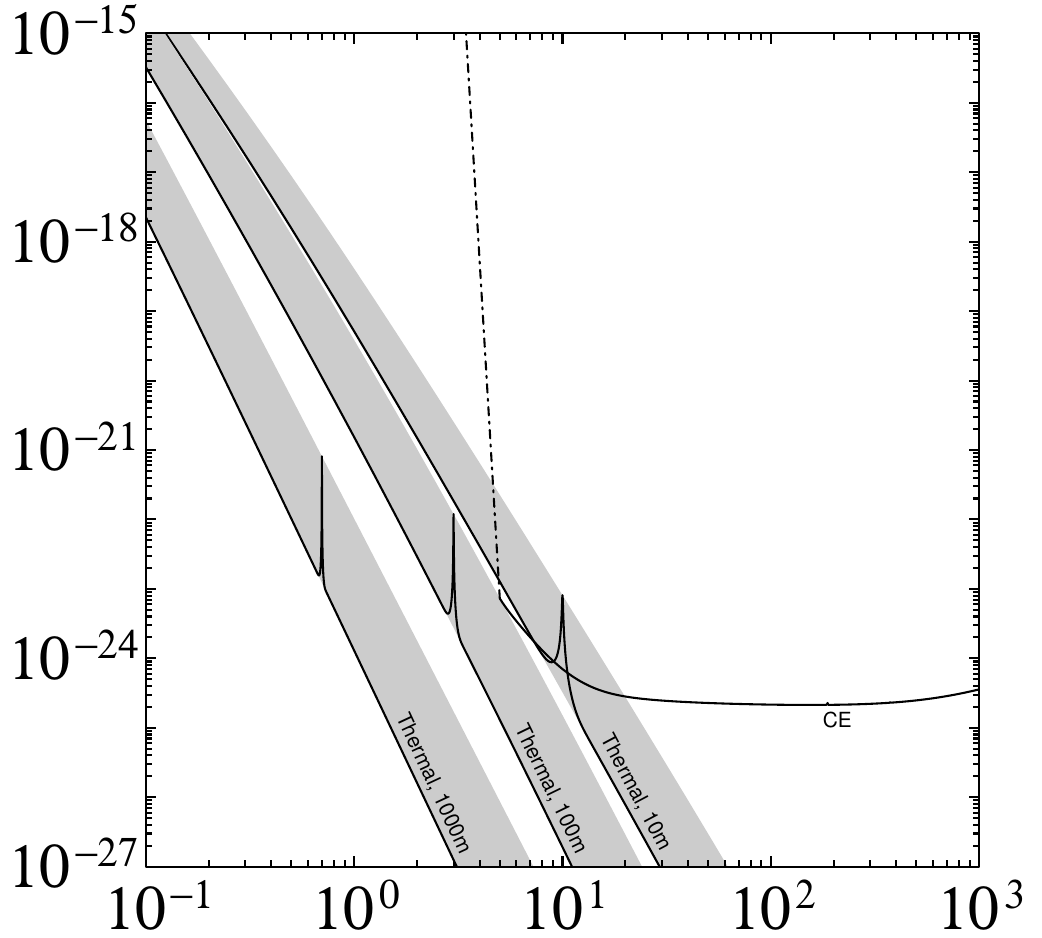}} \\
      (a) & $f$ (Hz) & (b) & $f$ (Hz)
    \end{tabular*}
  \end{center}
  \vspace{-2ex}
  \caption{\label{fig:ce} Estimated Newtonian noise levels in Cosmic Explorer (CE) from (a) infrasound and (b) thermal
    perturbations.  In each case, three noise estimates are plotted,
    assuming no perturbations within 10\,m, 100\,m, or 1000\,m of the
    test masses, respectively.  The line labeled CE shows the detector
    design noise from~\cite{ce-sensitivities} (dot-dashed lines are the authors' extrapolation). For infrasound, the solid curves correspond to highly coherent infrasound, while the tops of the shaded regions assume coherence lengths comparable to wavelength.  For thermal Newtonian noise, the bottoms of the shaded
    bands correspond to a broadband spectrum of turbulence, while the
    tops of the shaded bands are the loci of ``worst
    case'' noise levels for turbulence concentrated at a particular
    frequency: examples of such spectra are shown at arbitrarily
    chosen peak frequencies.}
\end{figure}

In Fig.~\ref{fig:ce}, we plot the infrasound and thermal noise spectra
against the Cosmic Explorer (CE) design sensitivity curve
from~\cite{et-sensitivities}, using the design length $L=40\,\mathrm{km}$.  Proposals have generally focused on surface-sited detector ($D\sim10\,\mathrm{m}$), but a subterranean detector has also been considered.  Here we give noise curves assuming
perturbations are excluded within 10\,m, 100\,m, and 1\,km of the test
masses.

Due to its longer arms and less-ambitious low-frequency response, Cosmic Explorer is somewhat less sensitive than Einstein Telescope to infrasonic Newtonian noise, as seen in figure~\ref{fig:et}(a).  However, if CE is located on the surface, it is still plausible that thermal Newtonian
noise would pessimistically exceed the design sensitivity at 5--10\,$\mathrm{Hz}$,
as shown in figure~\ref{fig:et}(b).  Proposals for surface vs.\ underground detectors should take this into consideration.

\subsection{Gradiometers}

\begin{figure}
  \begin{center}
    \begin{tabular*}{\hsize}{@{}c@{}c@{\extracolsep{\fill}}c@{\extracolsep{0pt}}c@{}}
      \rotatebox{90}{\makebox[2in][c]{$\sqrt{S_h}$ ($\mathrm{Hz}^{-1/2}$)}} &
      \resizebox{!}{2in}{\includegraphics{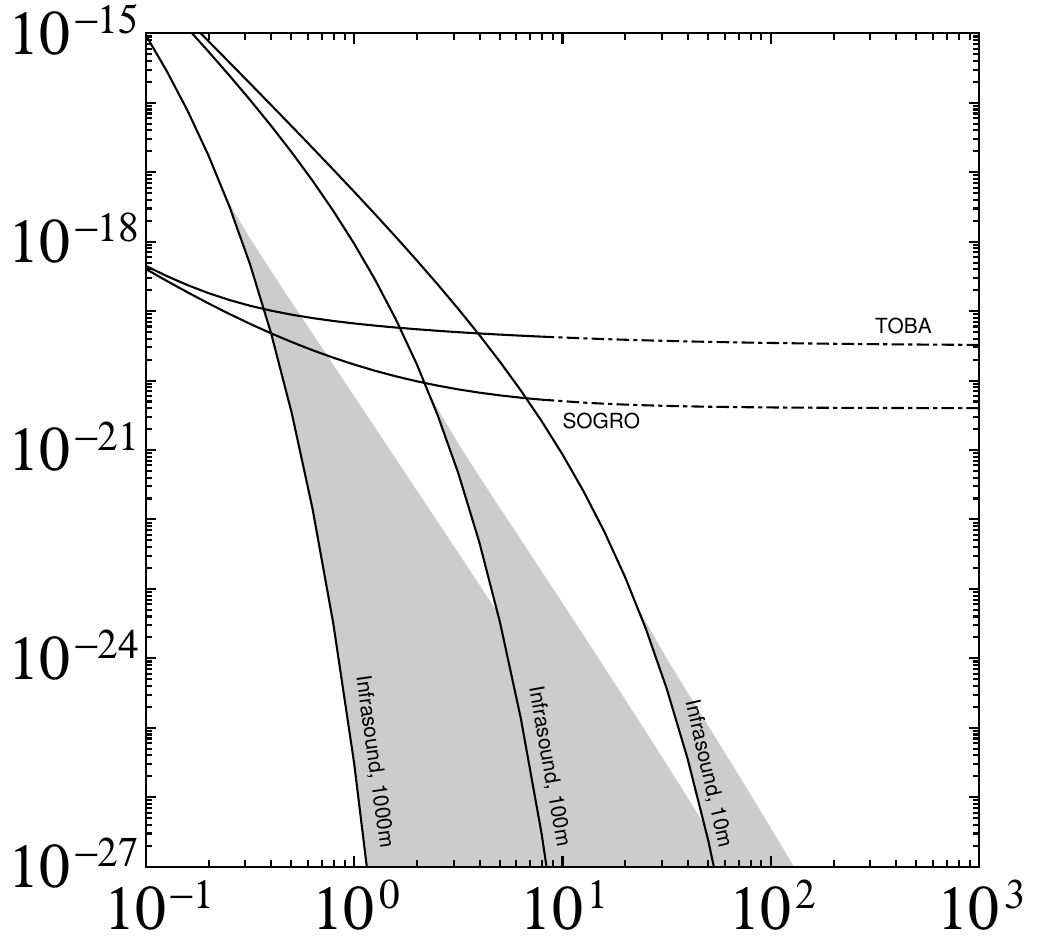}} &
      \rotatebox{90}{\makebox[2in][c]{$\sqrt{S_h}$ ($\mathrm{Hz}^{-1/2}$)}} &
      \resizebox{!}{2in}{\includegraphics{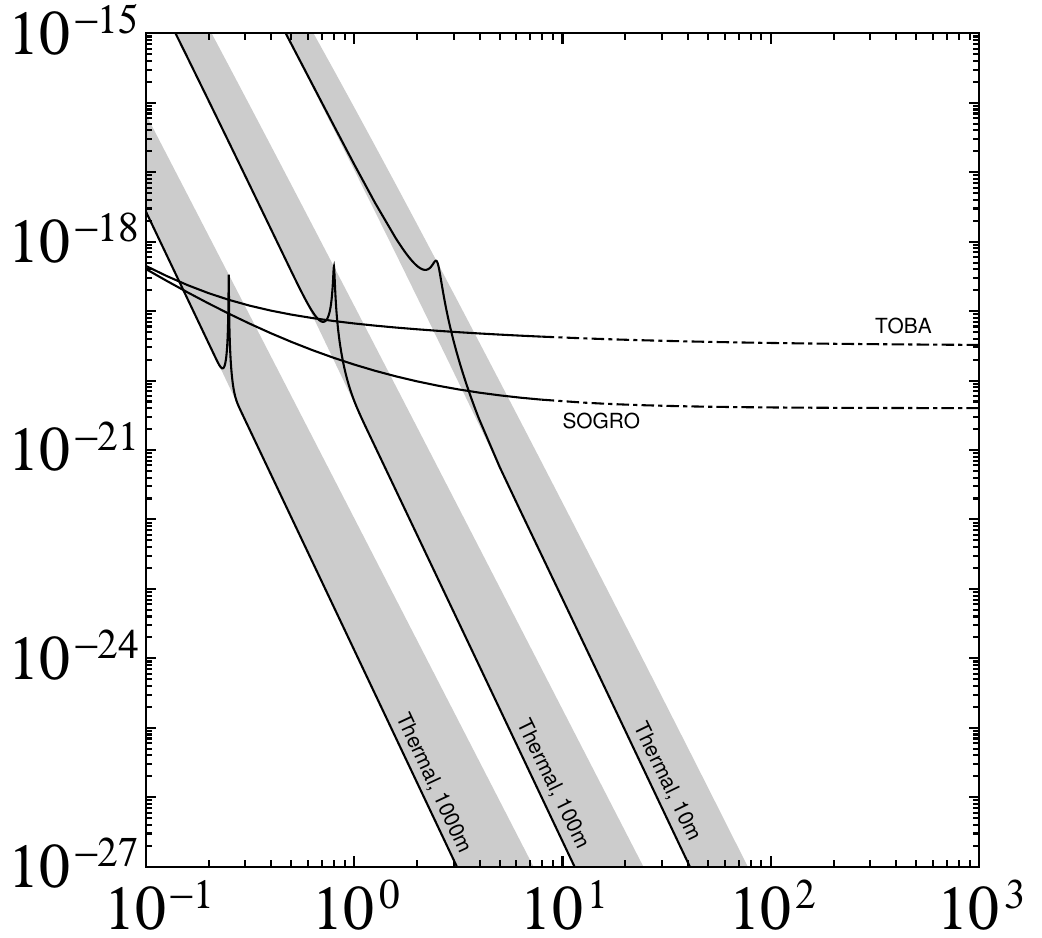}} \\
      (a) & $f$ (Hz) & (b) & $f$ (Hz)
    \end{tabular*}
  \end{center}
  \vspace{-2ex}
  \caption{\label{fig:gradiometer} Estimated Newtonian noise levels in
    gradiometric gravitational-wave detectors from (a) infrasound and
    (b) thermal perturbations.  In each case, three noise estimates
    are plotted, assuming no perturbations within 10\,m, 100\,m, or
    1000\,m of the test masses, respectively.  The lines labeled SOGRO and TOBA
    show detector design noises for two proposed detectors as given in~\cite{sogro:2016} and~\cite{toba:2020} respectively (dot-dashed lines are the authors' extrapolation).  For infrasound, the solid curves correspond to highly coherent infrasound, while the tops of the shaded regions assume coherence lengths comparable to wavelength.  For thermal
    Newtonian noise, the bottoms of the shaded bands correspond to a
    broadband spectrum of turbulence, while the tops of the shaded
    bands are the loci of ``worst case'' noise levels
    for turbulence concentrated at a particular frequency: examples of
    such spectra are shown at arbitrarily chosen peak frequencies.}
\end{figure}

In Fig.~\ref{fig:gradiometer}, we plot the infrasound and thermal noise
spectra against the design sensitivity curves for two gradiometric
gravitational-wave antennas: the Torsion Bar Antenna (TOBA)~\cite{toba:2020}
and the Superconducting Omnidirectional Gravitational Radiation
Observatory (SOGRO)~\cite{sogro:2016}.  As with the Einstein Telescope, one
would expect the final detector to be installed underground to
mitigate Newtonian noise: we present curves assuming depths of 10\,m
(i.e.\ within a building on the surface), 100\,m, and 1\,km.

From these plots, it is apparent that Newtonian noise sources are
likely to determine the low-frequency cutoff of these instruments.
Some gradiometric detectors are proposed to be ``full-tensor'' detectors, offering some discrimination between gravitational waves and Newtonian fields by measuring all components of the gravity gradient tensor~\cite{harms:2015}.  However, such discrimination still requires some independent measurement of the perturbing mass field, which is a challenge for atmospheric NN (and thermal NN in particular).  Pushing to sub-hertz frequencies will almost certainly require a
detector that is hundreds of metres away from any atmospheric
turbulence (for example, the Kamioka mine used to house
KAGRA~\cite{kagra:2018}), and even with such measures, sensitivity below $10^{-1}\mathrm{Hz}$ is dubious.

\subsection{Space Detectors}

These results can be applied to space detectors but only for those in low- to medium-Earth orbit, such as B-DECIGO~\cite{decigo:2017}.  Even then, we would expect the predominant Newtonian noise source to be seismic: given that there is no obvious way to model and subtract seismic noise over 1000-kilometre scales, we would expect it always to dominate over atmospheric NN.  For space detectors in Earth (rather than Solar) orbit, seismic NN will likely be a concern at low frequencies, but such an analysis is outside the scope of the current paper.

\section{Discussion}
\label{s:discussion}

This paper reproduces the analysis in~\cite{creighton:2008} in a more
general form, applicable to a broader range of interferometer designs.
In the specific case of advanced LIGO, it confirms the result
in~\cite{creighton:2008}, that atmospheric Newtonian noise will affect
the advanced LIGO noise floor only in the most pessimistic
circumstances.  However, for third-generation detectors, the
predictions are more serious: atmospheric Newtonian noise becomes the
principal constraint on extending the detection of gravitational waves
to lower frequencies.  A detector targeting frequencies below 10\,Hz
will likely need to be located hundreds of metres below ground, and
one aiming below 1\,Hz should be in Solar orbit.

Some effort has gone into planning and designing feed-forward noise
cancellation or noise subtraction schemes for seismic Newtonian noise,
and similar techniques may be applicable to coherent infrasound:
see~\cite{harms:2019} for a review.  However, little has been proposed
to mitigate wind-advected thermal Newtonian noise, which may prove to
be intractable on the timeframe of third generation detectors.

Both seismic and infrasonic noise cancellation efforts benefit from
the spatial coherence of these disturbances, allowing the effect of
the disturbance to be computed accurately from a sparse network of
seismic or infrasonic detectors.  Thermal perturbations, by contrast,
are presumed to be incoherent on lengthscales larger than the
disturbance itself.  For Newtonian noise at $\sim1$\,Hz at wind speeds
of $\sim10$\,m/s, the relevant lengthscale is $v/\omega\sim2$\,m; and to
achieve an effect similar to placing the detector a distance $D$ from
such disturbances, one must accurately model and subtract the effect
of such disturbances within a volume $\sim D^3$.  As an example example, for a
detector on the surface to achieve the same Newtonian noise at 1\,Hz
as a detector 100\,m underground, one would need an accurate snapshot
of $\sim10^6$ independent temperature measurements to millikelvin precision of metre-scale air
packets within 100\,m of the detector in three dimensions.

Developing the sensor array needed to perform such measurements, and
the computational infrastructure to apply the modeling and
subtraction, will be a significant task.  At first glance, \textit{in
  situ} measurements would seem to be impractical, making remote
sensing the only option.  Conceivably this could involve thermograph imagers sensitive to an atmospheric absorption line, or a lidar system that could measure variations in the index of refraction, using tomography to build up a three-dimensional density map.  Whether such a system could achieve the necessary
millikelvin, sub-second, and metre-scale precision remains to be seen.

The more straightforward solution is to physically restrict the airflow
and thermal environment near the detector.  Large inflatable above-ground structures are one possibility, but underground placement has the additional advantage of reducing seismic Newtonian noise.  This reinforces the
conclusion that third-generation detectors should be sited as far
underground as possible, if they are to achieve
sub-hertz sensitivity.

\section{Acknowledgments}

The authors thank Brian Lantz and Volker Quetschke for helpful discussions.  This work was supported by the National Science Foundation through grants NSF-1912630 and NSF-2207999.

\appendix
\section{Numerical Simulations of Thermal Newtonian Noise}

Thermal atmospheric Newtonian noise is potentially the largest immitigable noise source at low frequencies.  It is therefore important to have confidence in the validity of our analytic estimates.  In order to check our analytic calculations, we have performed numerical simulations of thermal Newtonian noise for several characteristic cases.

\begin{figure}
  \begin{center}
    \begin{tabular*}{\hsize}{@{}c@{}c@{\extracolsep{\fill}}c@{\extracolsep{0pt}}c@{}}
      \rotatebox{90}{\makebox[2in][c]{$\sqrt{S_h}$ ($\mathrm{Hz}^{-1/2}$)}} &
      \resizebox{!}{2in}{\includegraphics{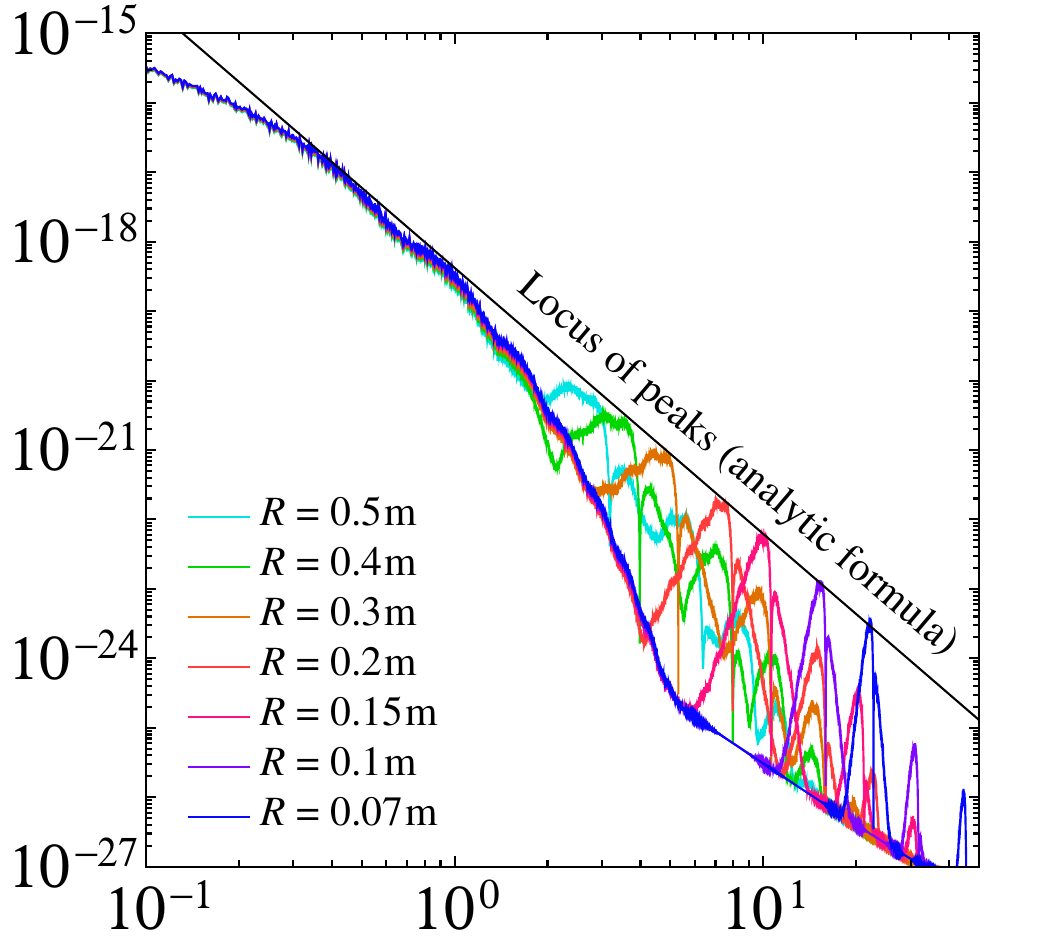}} &
      \rotatebox{90}{\makebox[2in][c]{$\sqrt{S_h}$ ($\mathrm{Hz}^{-1/2}$)}} &
      \resizebox{!}{2in}{\includegraphics{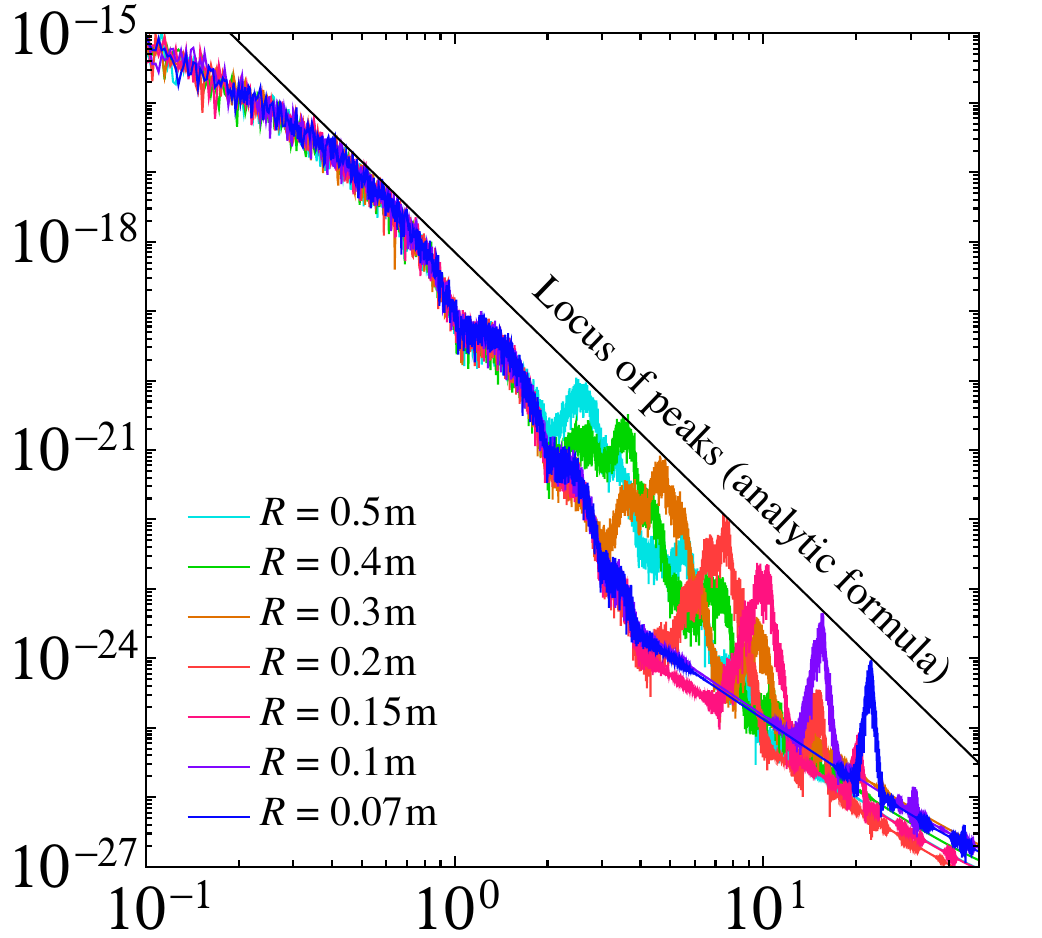}} \\
      (a) & $f$ (Hz) & (b) & $f$ (Hz)
    \end{tabular*}
  \end{center}
  \vspace{-2ex}
  \caption{\label{fig:numerical} Numerical simulation of thermal Newtonian noise from turbulent (cycloidal) airflow, for (a) 1-dimensional and (b) 2-dimensional models of the air mass.  Wind speed is $v=10\,\mathrm{m}/\mathrm{s}$.  Coloured lines represent simulations with different cycloid radii $R$, resulting in noise peaked at frequencies $f=v/2\pi R$.  The solid lines represent the analytic estimate of the locus of peaks, given by Eqs.~(\ref{eq:sh-peak-1d}) and~(\ref{eq:sh-peak-2d}), the 1d and 2d analogues (respectively) of Eq.~(\ref{eq:sh-peak}).  This indicates good agreement between analytic and numerical calculations, and validates the analytic model.}
\end{figure}

\subsection{1d Simulations}

We begin with simple one-dimensional simulations, where we restrict ourselves to linear density perturbations $\lambda(x)$ along a single streamline passing over the detector.  In order to compare with the formulae in section~\ref{s:thermal} we write $\lambda=\rho\delta A$, where $\delta A$ is the notional ``cross-section'' of the streamline; however, since the 1d numerical model does not actually impose a physical thickness to the streamline, it is more accurate to view $\delta A$ simply as a unit conversion factor.  It is fairly straightforward to adjust the derivation in~\cite{creighton:2008} to such a case: replacing the volume integrals integral in Eqs.~(A.1) of~\cite{creighton:2008} to one-dimensional integrals along the streamline.  The analogue to Eq.~(\ref{eq:sh-general}) is:
\begin{equation}
S_h(f) \;\;=\;\; \frac{G^2\rho^2\delta A^2}{L^2}\frac{c_T^2}{T^2}\omega^{-(p+5)}A(p)\sum_k \tilde{F}_{S,k}(f)^*\tilde{G}_{S,k}(f) v
\end{equation}
where $G(t)=x(t)/r^3(t)$ as before, and $F(t)=G(t)v(t)^{p+1}$ (note the different $v$ dependence in $F$).  There is no integral over streamlines, as the 1d case assumes a \emph{single} streamline $S$.  For a cycloidal streamline of radius $R=v/\Omega$ passing a distance $D$ above the detector, taking $v$ to be constant and evaluating at the peak of the spectrum $\omega=\Omega$, we obtain the 1d analogue of Eq.~(\ref{eq:sh-peak}):
\begin{equation}
  \label{eq:sh-peak-1d}
  S_h \;\;\lessim\;\; \frac{2\cos^2\theta G^2\rho^2\delta A^2}{D^4L^2}\,\frac{c_T^2}{T^2}\,
  \frac{v^{p+2}}{\omega^{p+7}}A(p)
\end{equation}

This is the formula that we tested numerically.  A random linear density field was generated by placing point masses along a line with a spacing 0.1\,m over a length 3000\,m.  The masses were initially generated as independent Gaussian-distributed random deviates, but the were then convolved with a spatial kernel to give the desired temperature/density correlations.

The \emph{temperature structure function} introduced in Sec.~\ref{s:thermal} is formally defined, in a homogeneous isotropic medium, as:
\begin{equation}
    D_T(|\!|\Delta r|\!|) \;\equiv\; \langle[T(\bb{r})-T(\bb{r}+\Delta\bb{r})]^2\rangle
\end{equation}
and is related to the temperature spatial correlation function:
\begin{equation}
    \langle T(\bb{r})T(\bb{r}+\Delta\bb{r})\rangle \;=\; \sigma_T^2 - \mbox{$\frac{1}{2}$}D_T(|\!|\Delta\bb{r}|\!|) \;.
\end{equation}
We have stated that on small scales, the structure function should go as a power law $D_T(\Delta r)=c_T(\Delta r)^p$, where $p=2/3$ is typical of Kolmogorov turbulence.  However, it is implicitly assumed that the correlation goes to zero at sufficiently large scales, so that $D_T$ saturates at some level (allowing us to write the temperature variance as $\sigma_T^2=\frac{1}{2}D_T(\infty)$).  For purposes of our simulation, our goal was to ensure that this saturation did not affect our results at the frequencies of interest, and to check that this was a reasonable assumption (in the absence of more detailed site-specific measurements and modeling).  Specifically, we sought to create a linear mass distribution of the form:
\begin{equation}
\label{eq:correlation-function}
    \langle\lambda(x)\lambda(x+\Delta x)\rangle \;\propto\; \langle T(x)T(x+\Delta x)\rangle \;\propto\; 1 - \left(\frac{\Delta x}{10\,\mathrm{m}}\right)^p
\end{equation}
where the choice of a 10\,m cutoff is both physically plausible, and will not significantly affect our result for frequencies above 1\,Hz.  (While in reality the turbulent boundary layer may extend hundreds of metres, modeling out to these distances would only affect our numerical results at the sub-hertz frequency scale.)

\begin{figure}
  \begin{center}
    \begin{tabular*}{\hsize}{@{}c@{}c@{\extracolsep{\fill}}c@{\extracolsep{0pt}}c@{}}
      \rotatebox{90}{\makebox[2in][c]{$D_T$ ($\mathrm{K}^2$)}} &
      \resizebox{!}{2in}{\includegraphics{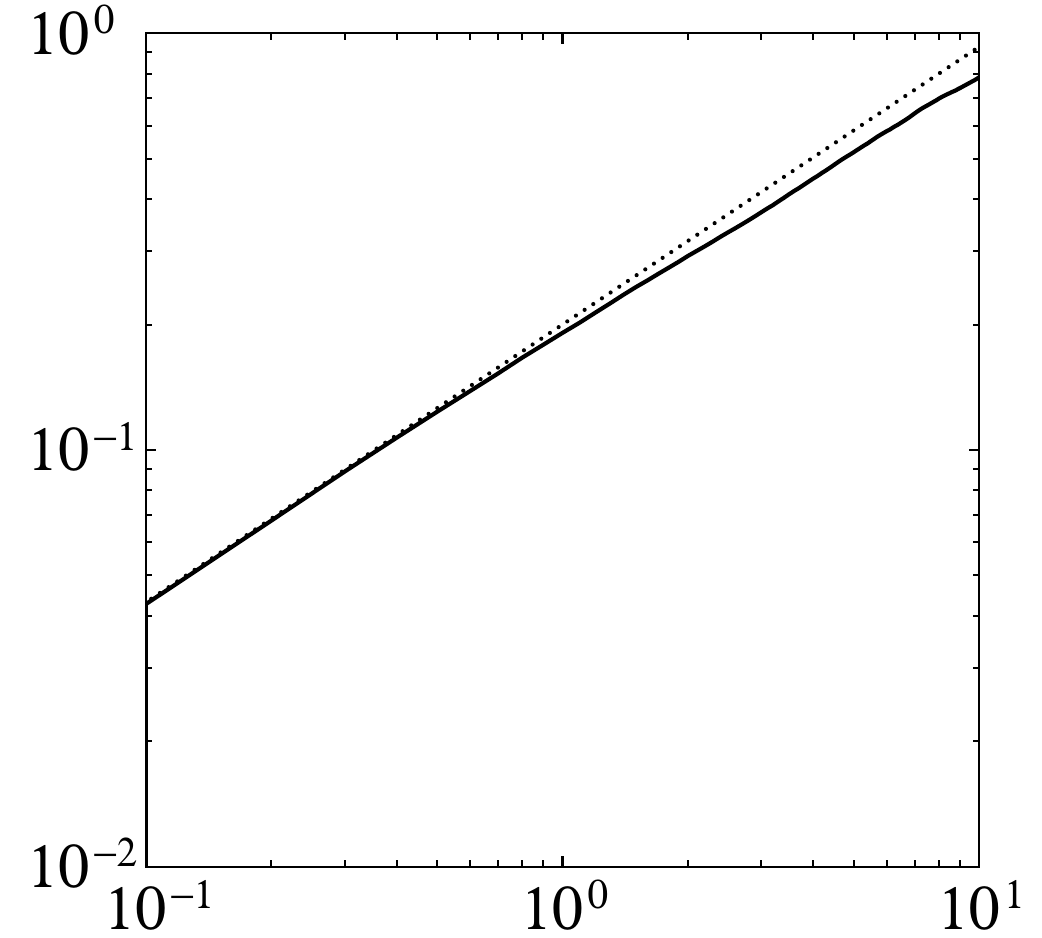}} &
      \rotatebox{90}{\makebox[2in][c]{$D_T$ ($\mathrm{K^2}$)}} &
      \resizebox{!}{2in}{\includegraphics{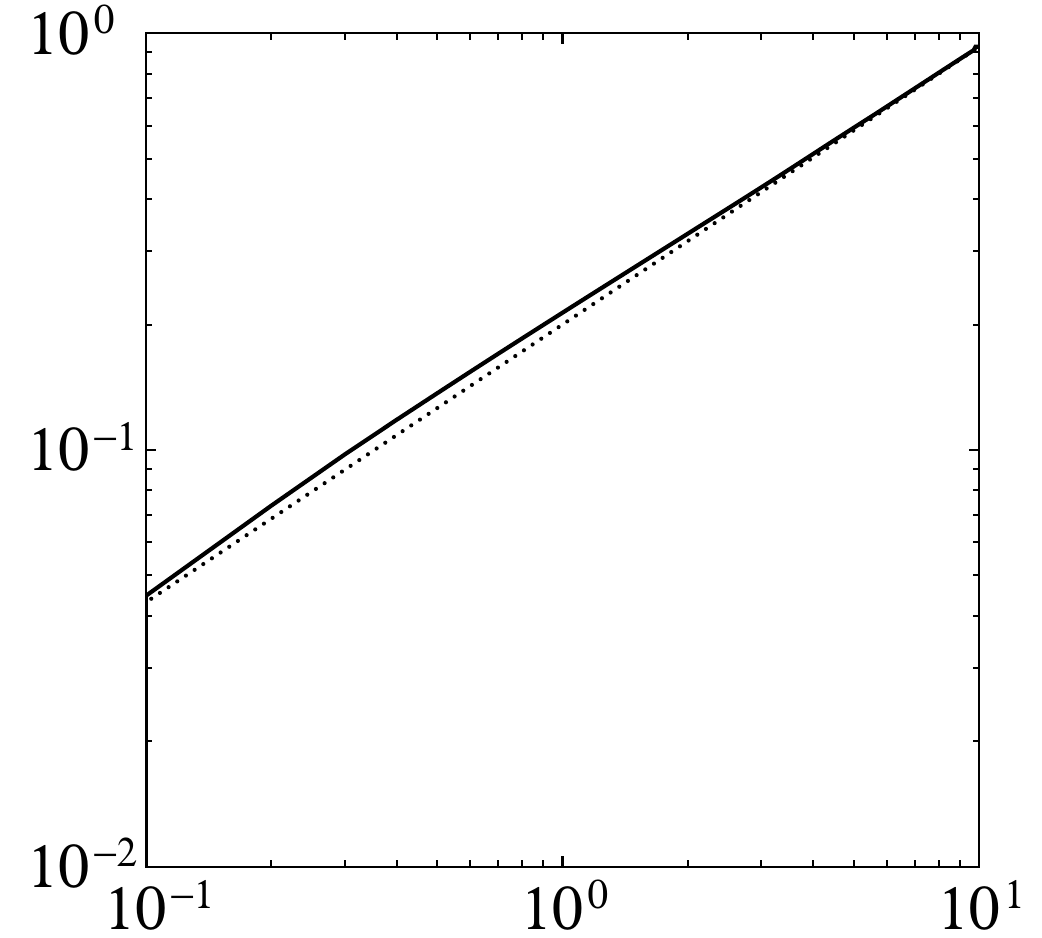}} \\
      (a) & $\Delta r$ (m) & (b) & $\Delta r$ (m)
    \end{tabular*}
  \end{center}
  \vspace{-2ex}
  \caption{\label{fig:structure} Temperature structure function $D_T(\Delta r)$ used in (a) 1-dimensional and (b) 2-dimensional models of the air mass.  Dotted lines are the theoretical structure function used in the analytic model: $D_T=c_T(\Delta r)^p$ with $p=2/3$ and $c_T=0.2\,\mathrm{K}^2m^{-2/3}$.  Solid curves are the actual structure functions of the air masses used in the numerical simulations, generated by convolving Gaussian temperature noise with a spatial kernel.  Within reasonable tolerances the numerical model follows the desired power law, at least for small distances (corresponding to high frequencies).}
\end{figure}

Convolution of uncorrelated white spatial noise with a kernel $K(x)$ results in a correlation function $\propto|K|^2$, so we chose a kernel equal to the square root of Eq.~\ref{eq:correlation-function}.  The overall proportionality was not specified in advance; instead, after performing the convolution, the structure function of the simulated airmass was measured, and density perturbations rescaled to match our assumed $c_T^2$.  At the same time we verified that the structure function had the correct $p=2/3$ power law.  This confirmatory measurement is shown in Figure~\ref{fig:structure}(a), where we have adopted a scaling factor $\delta A=1\,\mathrm{m}^2$ to relate 3d volumetric air densities to 1d linear densities.

The point masses were then translated along a cycloidal streamline with a cycloid radius $R$, at a speed $v=10$\,m/s at an angle $\theta=40^\circ$ to the detector arm, passing a distance $D=4\,\mathrm{m}$ above the test mass.  Positions were sampled at time intervals $\Delta t=0.01\,\mathrm{s}$ over $T=400\,\mathrm{s}$.  The spatial separation of the point masses and the time intervals were chosen so that sampling artifacts would not affect the results in a frequency range $\gtrsim0.005$\,Hz and $\lessim50$\,Hz.
At each sample time, we computed the net gravitational acceleration on the test mass in the direction of the arm.  The resulting time series {$\delta g(t)$} was then converted into a noise power spectral density $S_h(f)$ via:
$$
S_h(f)=\frac{1}{L^2(2\pi f)^4}\frac{1}{T}\left|\delta\tilde{g}(f)\right|^2\;,
\qquad \delta\tilde{g}(f_j)=\sum_k e^{2\pi ijk}\delta g(t_k)\Delta t\;.
$$
This was repeated for various cycloid radii $R$.  The resulting $S_h$ and the analytic prediction from Eq.~\ref{eq:sh-peak-1d} are shown in Figure~\ref{fig:numerical}(a).  We note the excellent agreement between the predicted locus and the computed noise peaks: not only do the peaks follow the expected power law, but also no fine-tuning of the overall amplitude factor was required.

\subsection{2d Simulations}

Encouraged by the results of the 1d simulations, we have extended our numerical validations to two dimensions.  In this case, the airmass is a sheet with some surface density $\sigma(x,y)=\rho\delta w$, where the dimensional conversion factor $\delta w$ can be thought of as the notional ``thickness'' of the sheet.  The analogue to Eq.~(\ref{eq:sh-general}) is:
\begin{equation}
S_h(f) \;\;=\;\; \frac{G^2\rho^2\delta w^2}{L^2}\frac{c_T^2}{T^2}\omega^{-(p+6)}B(p)\sum_k \int_{\{l\}}\tilde{F}_{S,k}(f)^*\tilde{G}_{S,k}(f) v_0\,dl
\end{equation}
where now $B(p)=2^{(p+5/2)}\sin(p\pi/2)\Gamma(p/2+1)^2$, $F(t)=G(t)v(t)^{p+2}$, and the integral $\int_{\{l\}}\ldots{}dl$ is the integral over streamlines in the sheet.  Again we assume constant $v$, cycloidal streamlines of radius $R=v/\Omega$, place the sheet a distance $D$ above the detector, and evaluate at $\omega=\Omega$, to give the two-dimensional analogue of Eq.~(\ref{eq:sh-peak}):

\begin{equation}
  \label{eq:sh-peak-2d}
  S_h \;\;\lessim\;\; \frac{\pi\cos^2\theta}{2}\frac{G^2\rho^2\delta w^2}{D^3L^2}\,\frac{c_T^2}{T^2}\,
  \frac{v^{p+3}}{\omega^{p+8}}B(p)
\end{equation}

For the numerical simulations, once again we started with point masses evenly spaced along the plane with separation $0.1$\,m in both axes.  The mass-filled area was 3000\,m long (to allow for a sufficiently long integration as it passed over the test mass), but only 40\,m wide (still wide enough to ensure that edge effects were an order of magnitude less than the effect of air passing directly over the test mass).  Masses were drawn from a Gaussian random deviate, and then convolved with the same spatial kernel as previously (i.e.\ the square root of Eq.~\ref{eq:correlation-function}).  In this case the convolution was done in 2 dimensions by means of Fourier transforms.  That is, a numerical 2d Fourier transform was performed on both the mass distribution and the kernel, the resulting wavevector data multiplied, and then inverse transformed back to the spatial domain.  As in the 1d case, the validity of this method was checked by measuring the structure function, and verifying its power law, as shown in Figure~\ref{fig:structure}(b).

The air masses were then translated along cycloidal streamlines of radius $R$, flowing in the $x$ direction with vertical displacements in the $z$ direction.  We note that in the 2d case, we must consider spatial correlation not only of the mass perturbations, but of the airflow pattern as well.  Our simulation followed the worst-case assumptions of Sec.~\ref{ss:smooth-vs-turbulent-airflow}: all streamlines were given the same cycloid radius $R$ (and hence the same frequency $f=v/2\pi R$).  Furthermore, nearby cycloids were assumed to be spatially correlated on a scale $\sim R$.  This was achieved by giving each streamline (characterized by its $y$-coordinate) an initial uniformly-distributed random cycloid phase $\phi(y)$.  This function was then numerically convolved with a Gaussian kernel of width $R$, so that cycloids within this characteristic distance would be progressively more ``in phase'' with one another.

As in the 1d case, the gravitational acceleration $\delta g(t)$ on the test mass was computed at each time, then converted to a noise power spectral density $S_h$.  This was then compared to the analytic prediction.  The results are shown in Figure~\ref{fig:numerical}(b).  We have used a dimensional scaling factor $\delta w=1\,\mathrm{m}$ to relate volumetric and surface densities.

Once again there is good agreement both in the power law of the locus and in overall amplitude.  We conclude that the analytic formulae are substantially correct, and give reliable estimates of the Newtonian noise for the cases of greatest concern.

\subsection{3d Simulations}

A full three-dimensional simulation is substantially more challenging both in algorithm design and computational cost.  We expect that, at this level of complexity, a full hydrodynamic simulation would be more fruitful than the current approximation of advection along fixed cycloidal streamlines.  Such a simulation is beyond the scope of the current paper and will be left to future work.

\section*{References}

\bibliography{main}{}

\end{document}